\documentclass[pdflatex,sn-mathphys]{sn-jnl}

\usepackage{amsmath,amsfonts}
\usepackage{graphicx}
\usepackage{textcomp}
\usepackage{float}
\usepackage{multirow}
\usepackage{subfig}

\jyear{2021}%

\theoremstyle{thmstyleone}%
\theoremstyle{thmstyletwo}%

\theoremstyle{thmstylethree}%

\begin{document}

\title[Badri et al.]{ Efficient Placement and Migration Policies for an STT-RAM based Hybrid L1 Cache for Intermittently Powered Systems}


\author*[1]{\fnm{SatyaJaswanth} \sur{Badri}}\email{2018csz0002@iitrpr.ac.in}

\author[1]{\fnm{Mukesh} \sur{Saini}}\email{mukesh@iitrpr.ac.in}

\author[1]{\fnm{Neeraj} \sur{Goel}}\email{neeraj@iitrpr.ac.in}

\affil[1]{\orgdiv{Computer Science and Engineering}, \orgname{IIT Ropar}, \orgaddress{\city{Rupnagar}, \postcode{140001}, \state{Punjab}, \country{India}}}



\abstract{

The number of battery-powered devices is rapidly increasing due to the widespread use of IoT-enabled nodes in various fields. Energy harvesters, which help to power embedded devices, are a feasible alternative to replacing battery-powered devices. In a capacitor, the energy harvester stores enough energy to power up the embedded device and compute the task. This type of computation is referred to as intermittent computing. Energy harvesters are unable to supply continuous power to embedded devices. All registers and cache in conventional processors are volatile. We require a Non-Volatile Memory (NVM)-based Non-Volatile Processor (NVP) that can store registers and cache contents during a power failure. \\

NVM-based caches reduce system performance and consume more energy than SRAM-based caches. This paper proposes Efficient Placement and Migration policies for hybrid cache architecture that uses SRAM and STT-RAM at the first level cache. The proposed architecture includes cache block placement and migration policies to reduce the number of writes to STT-RAM. During a power failure, the backup strategy identifies and migrates the critical blocks from SRAM to STT-RAM. When compared to the baseline architecture, the proposed architecture reduces STT-RAM writes from 63.35\% to 35.93\%, resulting in a 32.85\% performance gain and a 23.42\% reduction in energy consumption. Our backup strategy reduces backup time by 34.46\% when compared to the baseline.

}


\maketitle

\section{Introduction}\label{sec1}

The Internet of things (IoT) has created several fascinating applications that consist of intelligent sensors and systems. IoT may consist of billions of sensors and systems by the end of 2050 \cite{big}. This prediction is exciting and promising, but deciding how to power these IoT devices is the main challenge. The majority of IoT devices are battery-powered. In some areas, such as deep mines, space, and industrial environments, replacing batteries after installation is difficult and expensive.

Furthermore, the battery has a specific problem with its lifetime  \cite{lifetime}. As a result, an alternative and the promising solution is to replace the battery with energy harvesters. Energy-harvesting devices extract energy from their surroundings, such as light, vibration, radio, and many others \cite{energy12}. The accumulated energy is used for powering up these IoT devices.

The unpredictable nature of energy harvesters causes voltage fluctuation or a power failure. A voltage stabilizer or a capacitor is a standard solution to the voltage fluctuation issue \cite{capac}. In a conventional processor, however, power failures result in data loss. The data is lost because registers, caches, and main memories are designed using volatile memories, such as SRAM and DRAM \cite{self}, \cite{power}. Data lost includes the application's program state and progress. The contents of registers, cache, and main memory are all part of the program state. As a result, when the power failure occurs, some parts of the application have to re-execute, causing the execution progress to be slow and consuming extra energy. This type of computing is known as intermittent computing \cite{intermittent,int2,int3}.

The solution is to store the application's program state at a precise restart point before a power failure. The question is, where should the program state be stored? Using Non-volatile memories (NVM), we can save the application's program state during a power failure. There are several new NVMs proposed recently, such as spin-transfer torque RAM (STT-RAM) \cite{jog, stt1, stt2, stt3}, phase-change memory (PCM) \cite{pcm}, resistive random-access memory (Re-RAM) \cite{r8}, and ferroelectric RAM (FRAM) \cite{msp}.

Researchers have explored NVM-based non-volatile processors (NVPs) \cite{wang}, which help in executing the application during the irregular power supply. NVM cache enhances the application's execution progress even during frequent power failures. NVMs have longer read and write latency than SRAM-based caches. Replacing SRAM with NVM is not a good idea; as an alternative, we can integrate both SRAM and NVM to make a hybrid architecture \cite{12} \cite{ma} at the cache level. Xie et al. \cite{12} propose a hybrid cache architecture for intermittently powered IoT devices.

This paper builds on previous work \cite{12} to improve the following aspects of a hybrid cache: (a) performance, (b) energy utilization, and (c) reducing writes to NVM.

Most of the literature mentions STT-RAM as an emerging candidate among all NVM technologies for LLC \cite{mao}, \cite{sun1}, \cite{smull}. STT-RAM promises higher density and less leakage power than existing SRAM. The main memory in conventional processors is implemented using DRAM, but the main memory in emerging micro-controllers, such as the MSP430FR5969 \cite{msp}, has non-volatile based main memory. For main memory, PCM is the appropriate memory technology because PCM has similar endurance features with STT-RAM and is also cheaper than other NVM technologies. As a result, throughout this paper, we have used STT-RAM as a non-volatile cache and PCM as a non-volatile main memory.

At the L1 cache, we propose Efficient Placement and Migration policies for hybrid cache architecture that includes both SRAM and STT-RAM. We assume that a capacitor's energy can backup the processor state during a power failure \cite{capacitor}. Because SRAM cache and SRAM-based register values are volatile, they must be stored in NVM. The proposed architecture identifies the blocks that need to be written to the main memory and STT-RAM at the L1 cache to reduce backup time.

Further, the exception mechanism of the pipelined processor is used to arrive at a precise wake-up point. When power comes back, our proposed architecture works like a regular architecture, i.e., every memory access is first looked up in the L1 cache. Due to the presence of STT-RAM in the L1 cache, it stores frequently accessed blocks, and the proposed architecture has benefits because of hybrid NVM cache architecture during frequent power failures.

In addition, we proposed a prediction table to help in block placement and migration. Compared to the baseline architecture, the proposed architecture improves performance by 32.85\% and reduces energy consumption by 23.42\%. Compared to the baseline, our proposed backup strategy reduces backup time by 27.91\%. The proposed architecture has a storage overhead of only 2.34\%.

This paper is organized as follows: Section \ref{p2} discusses the related works. Section \ref{system} explains the motivation behind the proposed system architecture and gives an overview of the problem formulation. Section \ref{p4} explains about proposed hybrid cache architecture. The experimental setup and results are described in section \ref{p5}. We concluded this work in section \ref{p6}.

\section{Related works} \label{p2}
This section reviews the related work in hybrid cache architectures (HCAs) and NVM for last-level caches (LLCs), HCA for L1 cache, and architectures for intermittent powered IoT devices. 

\subsection{\textbf{NVM for LLC} }


STT-RAM offers better features than the existing NVM technologies \cite{stt1,stt2,stt3}. In order to use STT-RAM at LLC, we have two possible and distinct preferences. First, by replacing the whole SRAM cache with STT-RAM at LLC. Second, using HCA (SRAM+STT-RAM) at LLC, where this design takes advantage of both SRAM and STT-RAM. 

Wu et al. \cite{wu} modeled a 3-level cache architecture by replacing SRAM with STT-RAM at the L3 and using SRAM at L1 and L2. This architecture achieved instructions per cycle (IPC) improvement of around 4\% and compared with the traditional 3-level SRAM cache design Wu et al. achieved a 63\% reduction in power consumption. 

Usually, for hybrid caches, block placement and movement between caches are the main challenges. Classifying the cache blocks based on write frequencies \cite{wang} and write access behavior \cite{5} \cite{7} \cite{prediction} \cite{10} that helps to decide where to place the respective cache block. Many architectures with the HCA use prediction table-based techniques to predict and place the cache block in an appropriate cache region and migrate from one cache to another \cite{prediction}. Challenges at L1 HCA are different from LLC. At LLC, input traffic is due to misses of L1/L2, while read/write requests at L1 are due to load/store instructions.

\subsection{\textbf{HCA for L1 caches}}

The write access latency of STT-RAM is higher than the SRAM, which creates the primary limiting factor of using STT-RAM at the L1 cache. For this concern, there are two possible alternatives. First, relaxing the non-volatility of STT-RAM to reduce the overall STT-RAM's write access latency \cite{sun1}, \cite{smull}. Relaxing the STT-RAM's non-volatility is achievable by reducing the MTJ planar area, and MTJ switching current \cite{energy}. Second, reducing the STT-RAM's write latency and energy consumption by limiting the number of writes to STT-RAM. Usually, the number of reads and write operations in the L1 cache is more than in the LLC. 

Xie et al. \cite{12} introduce an HCA that consists of STT-RAM in the L1 cache. During power failures, they backup the program state from the SRAM cache to the STT-RAM cache. The authors use an access pattern-based predictor that predicts block behavior. Based on the prediction, Xie et al. place the cache block in the respective cache region. During an eviction or on a wrong prediction, Xie et al. propose a migration policy that migrates a cache block from one cache region to another. Whenever power comes back, Xie et al. restore the cache contents from the STT-RAM cache to the SRAM cache. 

In these hybrid caches, wrongly placing a cache block in any region causes migration overhead. Migration overhead increases the number of writes to NVM and consumes more write energy. Introducing NVM in the L1 cache shows an impact on performance and energy consumption. Therefore, we proposed an efficient HCA to address the above issues. We use SRAM and STT-RAM at the L1 cache to reduce these migration overheads during both stable and unstable power scenarios. We proposed placement and migration policies, which also have a prediction table to predict the correct placement to reduce these additional overheads. 


\subsection{ \textbf{Architectures for Intermittent power devices}}

NVM-based NVPs \cite{fixing}, \cite{design}, \cite{ene} are proposed by storing the contents of the registers, volatile on-chip data to the non-volatile registers, and non-volatile memories, respectively. Whenever power comes back, the system uses data from the NVM region to continue and complete the application execution. 

Checkpoint-based approaches for HCA are the other alternative for supporting intermittently powered IoT devices. In these checkpoint-based approaches, volatile data is checkpointed to NVM at regular intervals to store the application program state \cite{checkpoint}. Mementos \cite{mementos} was one of the initial checkpointing techniques. It used periodic voltage checks to decide when to back up the program state. Hibernus \cite{hibernus++} extended the work of \cite{mementos} by introducing NVM. These checkpointing schemes don't consider the timely execution of the applications. TICS \cite{tics} overcomes this problem by introducing timely execution, branching, and efficient automatic checkpoints. 

Checkpoints are placed using either software procedures or hardware components. Checkpointing approaches like \cite{mementos}, \cite{ferroelectric}, and \cite{achieving} were proposed to backup and restore a consistent program state. The compiler or software procedures were primarily responsible for placing software-based checkpoints. Whenever a checkpoint is identified, the system initiates a backup procedure that stores the program state to NVM. In \cite{achieving}, checkpoints are placed based on the expiration of a timer. Hardware-based checkpoints were mainly associated with external devices. In \cite{ferroelectric}, hardware-based checkpoints were placed using a voltage detector that triggers a backup mechanism for an NVP. 

The main problem with the above architectures and techniques is the extra computation caused by multiple checkpoints. Another issue with checkpointing during a power failure is data inconsistency, which leads to a corrupted output. Another disadvantage of the checkpointing approach is that whenever power comes back, we must restore the contents of non-volatile main memory to the cache. Whenever power comes back, we must implement a restoration procedure that restores the saved checkpoint from the NVM. Restoring the program state introduces one more extra overhead. We proposed an HCA with a backup policy that triggers during a power failure instead of multiple backups of the program state at the desired checkpoints. Instead of restoring the program state after every power failure, our proposed HCA implements an automatic restoration process by accessing the data from the NVM.

\section{Motivation and Problem Formulation} \label{system}

This section discusses observations that motivate us to propose new architecture and techniques. We performed a set of experiments on a system configuration that consists of equal SRAM and STT-RAM at the L1 cache. In section \ref{p5}, we have given more details of the experimental setup, and table \ref{tab1} has architectural parameters.


\begin{figure}[htp]
    \centering
    \subfloat[\centering Execution Time ]{{\includegraphics[width=1\linewidth]{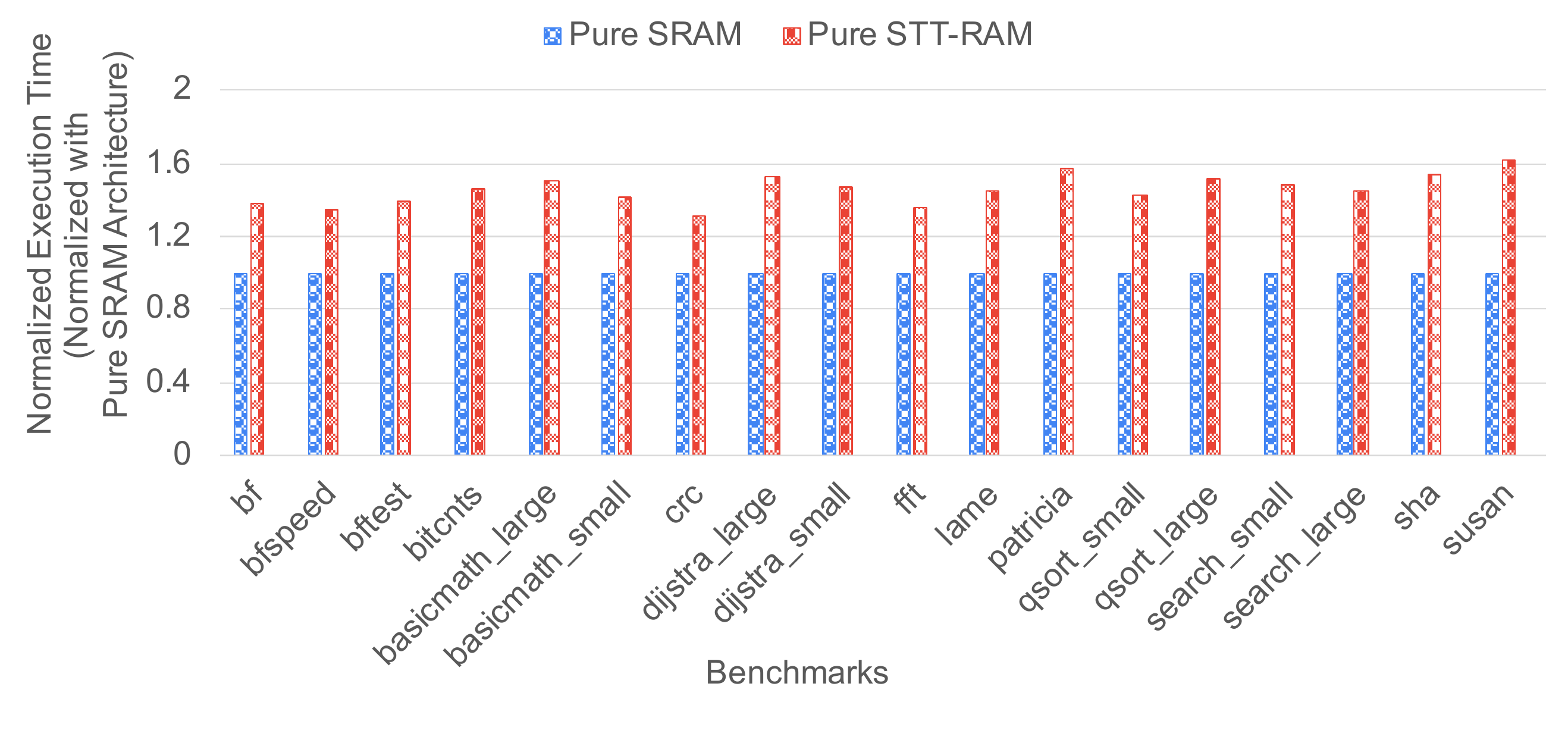} }}%
    \qquad
    \subfloat[\centering Dynamic Energy Consumption ]{{\includegraphics[width=1\linewidth]{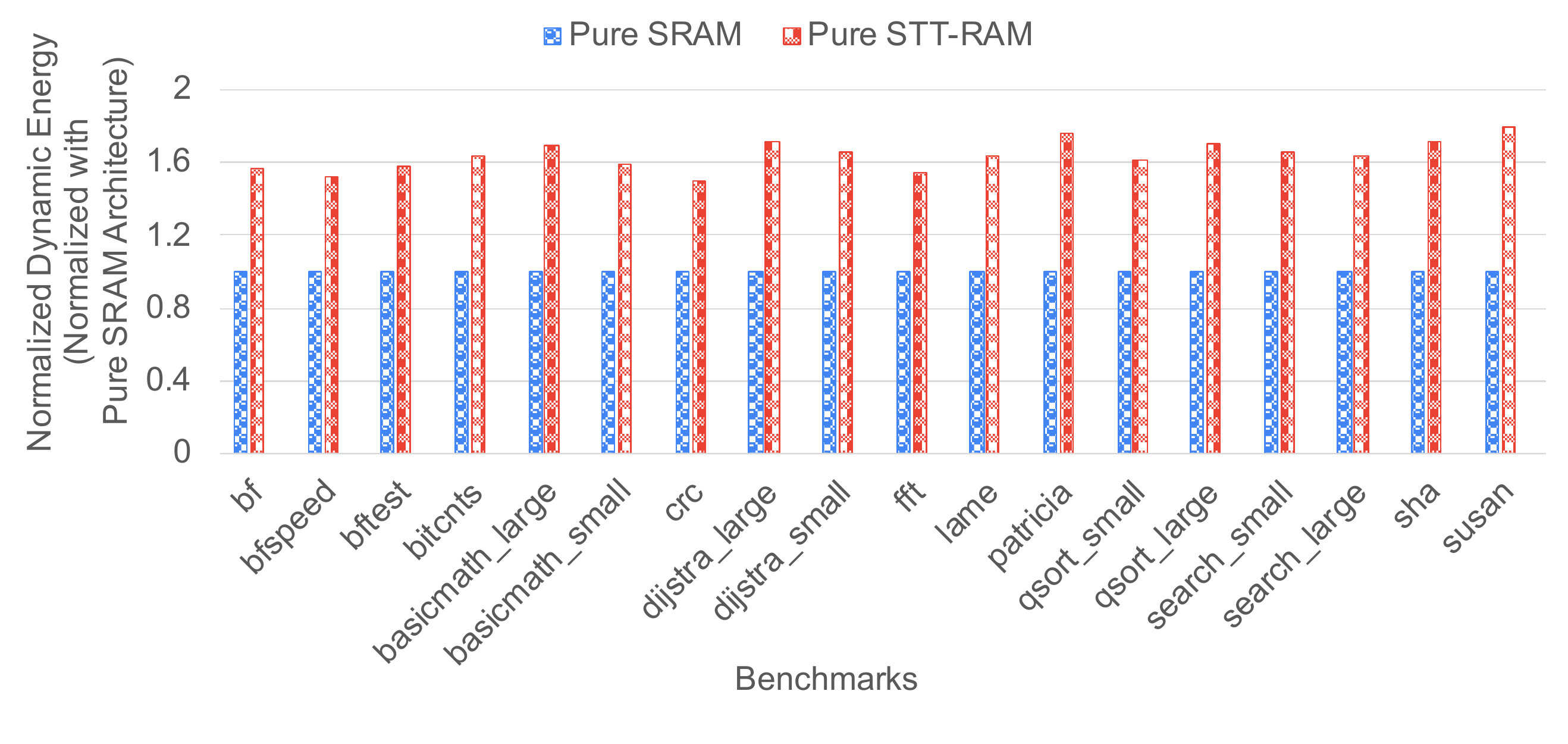} }}%
    \caption{Comparisons between Pure SRAM and Pure STT-RAM Architectures in terms of Execution time and Dynamic Energy Consumption}%
    \label{overhead1}%
\end{figure}

\subsection{Motivation}  \label{overhead}

Introducing STT-RAM as a cache can deteriorate the system's performance due to its long access time and consumes more dynamic energy. We modeled two cache architectures in gem5 \cite{13}, pure SRAM cache (only SRAM at L-1) and pure STT-RAM cache (only STT-RAM at L-1), to compare their performances and energy consumption.

In figure \ref{overhead1}, the performance and energy consumption of the cache architectures are normalized based on the pure SRAM cache architecture. Figure \ref{overhead1} (a) shows that STT-RAM cache architecture takes 45.93\% more execution time than pure SRAM cache architecture. Our first observation is to use STT-RAM efficiently so that it should not deteriorate the overall system performance and energy consumption. Thus, we need to use a hybrid cache instead of a pure STT-RAM cache, where that hybrid cache benefits from both SRAM and STT-RAM. 

In the case of HCAs, movement between two cache regions was explored in literature, i.e., migration-based policies for hybrid caches \cite{sun}, \cite{wuhybrid}, \cite{11stt}, \cite{high}, \cite{choi}. Migrating a cache block from one cache region to another cache region yields extra overheads, i.e., migration overheads. These overheads increase the number of reads and write operations and require additional cycles and energy, making the system inefficient by consuming more energy and deteriorating the overall system's performance. Thus, our second observation is to reduce these additional migration overheads.

The observations and challenges mentioned above motivated us to propose an HCA. We proposed an HCA that uses both SRAM and STT-RAM efficiently, with the proposed architecture that benefits from SRAM during regular operation and STT-RAM during power failure.

In the existing architectures, Xie et al. \cite{12} also introduced a similar hybrid cache architecture that consists of STT-RAM at the L1 cache. The main observations that we reported from the Xie et al. work and the main challenges associated with the existing HCA \cite{12} are listed below.

\begin{enumerate}
  
   \item For the prediction table, Xie et al. used a pattern sampler, which doesn't gather the complete details of the application.

\item Where the placement and migration policies cannot provide accurate predictions if the prediction information is incomplete. As previously stated, inaccurate predictions increase the number of reads and write operations, which consume more execution time and energy.

\item Xie et al. used a checkpointing scheme, which uses more energy because we need to write/read to/from the NVM for every checkpoint.

 \item Xie et al. used a standard LRU replacement policy to identify the cache block for eviction. What if the evicted block turns out to be a write-intensive block the next time? The used replacement policy may result in unnecessary writes to NVM and consumes more energy.

\item Xie et al. backup all volatile contents during a power failure, which is not always necessary, and push more writes to NVM during frequent power failures.

\end{enumerate}

All the above challenges and observations motivated us to propose an efficient HCA that considers these issues. In section \ref{p4}, we discussed the proposed HCA and placement policies in detail.








\subsection{Problem definition} \label{pdef}
\begin{figure}[htp]
\centering
\includegraphics[width= 0.40\linewidth]{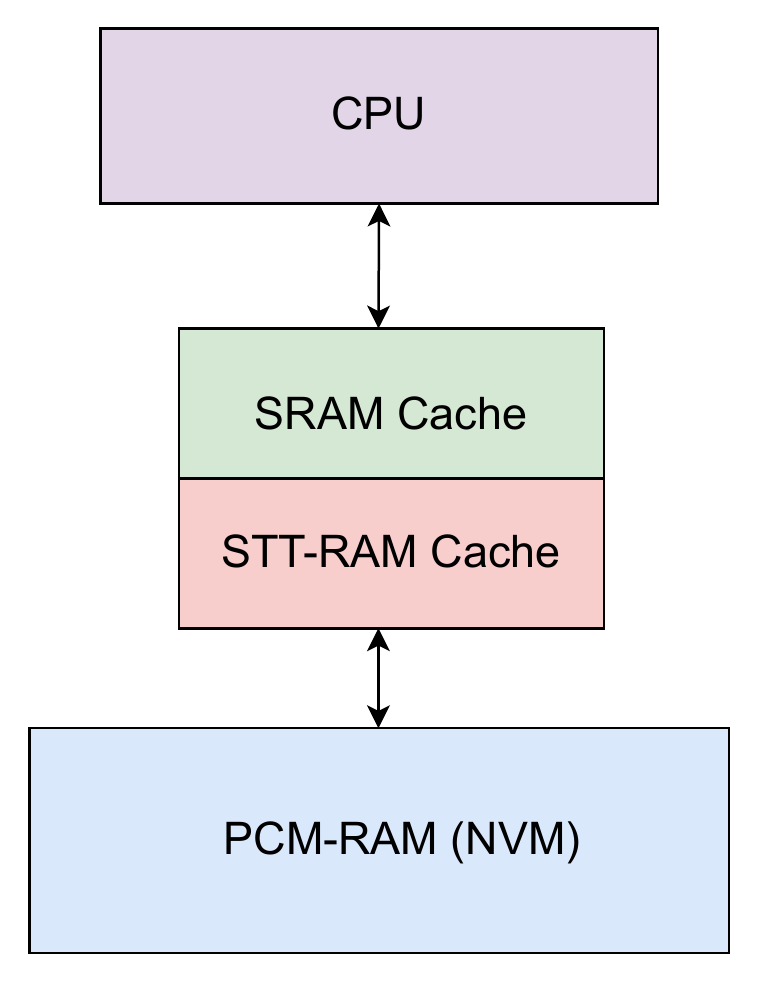}
\caption{Proposed System Model}
\label{stt1}
\end{figure}

We propose a hybrid cache model as shown in figure \ref{stt1}. By introducing NVM at the L1 cache, we observed additional overheads, which were discussed in section \ref{overhead}. So we reduce these overheads by introducing placement and migration policies. We formulated our three main objectives below.

\begin{itemize}
    \item Minimize backup energy.
    \item Maximize backup efficiency.
    \item Maximize energy efficiency.
\end{itemize}



For the hybrid architecture model, our design performs a backup during a power failure, and when the power comes back, it performs a memory restore operation. As a result, we define the energy required to execute the application in equation \ref{eq1}. 

\begin{equation} \label{eq1}
    E_{overall} = E_{exec} + E_{backup} + E_{restore}
\end{equation}

Where $E_{overall}$ is the energy required to execute the overall application, $E_{exec}$ is the energy required to execute the program. We backup the system state by copying all register contents and SRAM cache blocks to NVM. The energy consumed by the backup procedure is $E_{backup}$, where it depends on the number of bytes to be backed up to NVM as shown in equation \ref{eq3}.

\begin{equation}\label{eq3}
    E_{backup} = N_{w\_L1} * e_{w\_sttram} + N_{w\_main} * e_{w\_pcm}  
\end{equation}

Where $N_{w\_L1}$ is the number of writes to STT\_RAM, $N_{w\_main}$ is the number of writes to main memory, $e_{w\_sttram}$ is the energy per write for the STT-RAM and $e_{w\_pcm}$ is the energy per write for PCM RAM. We achieve our first objective by reducing the number of writes, $N_{writes} = N_{w\_L1} + N_{w\_main}$, during both stable power and intermittent power supply.

The energy required to restore the volatile contents from NVM is $E_{restore}$, where it depends on the number of bytes to be restored from NVM and can be defined as follows:

\begin{equation} \label{eq4}
    E_{restore} = N_{r\_L1} * e_{r\_sttram} + N_{r\_main} * e_{r\_pcm}  
\end{equation}

Where $N_{r\_L1}$ is the number of reads to STT\_RAM, $N_{r\_main}$ is the number of reads to main memory, $e_{r\_sttram}$ is the energy per read for the STT-RAM and $e_{r\_pcm}$ is the energy per read for PCM RAM. Whenever power comes back, the size of the restoring contents is the same as the content that was backed up during a power failure. Thus, the equations \ref{eq3} and \ref{eq4} were interrelated in terms of sizes, and as we are doing automatic restoration, so we don't have any restore overhead in our proposed HCA.


Our second objective is maximizing backup efficiency ($\eta$), defined and shown in equation \ref{eq10}.

\begin{equation} \label{eq10}
    \eta = \frac{N_{w\_L1}}{N_{w\_L1} + N_{w\_main}}
\end{equation}

If we achieve less $N_{writes}$, our $\eta$ increase. Thus, we achieve our second objective by reducing $N_{writes}$.

Lastly, we define energy efficiency as the ratio of energy consumed during normal execution without any power failures to the energy consumed during power failures. Let $\theta$ be the energy efficiency as defined in the equation \ref{eq11}. 

\begin{equation} \label{eq11}
\begin{split}
        \theta &= \frac{E_{normal}}{E_{overall}}
\end{split}
\end{equation}

Where $E_{normal}$ is the energy required for normal execution without any power interruptions.

\section{Proposed Architecture} \label{p4}
This section explains the proposed architecture that uses the proposed placement, migration, and backup policies.

\subsection{Hybrid Cache Architecture} \label{HCA}

\begin{figure}[htb]
\includegraphics[width= 1\linewidth]{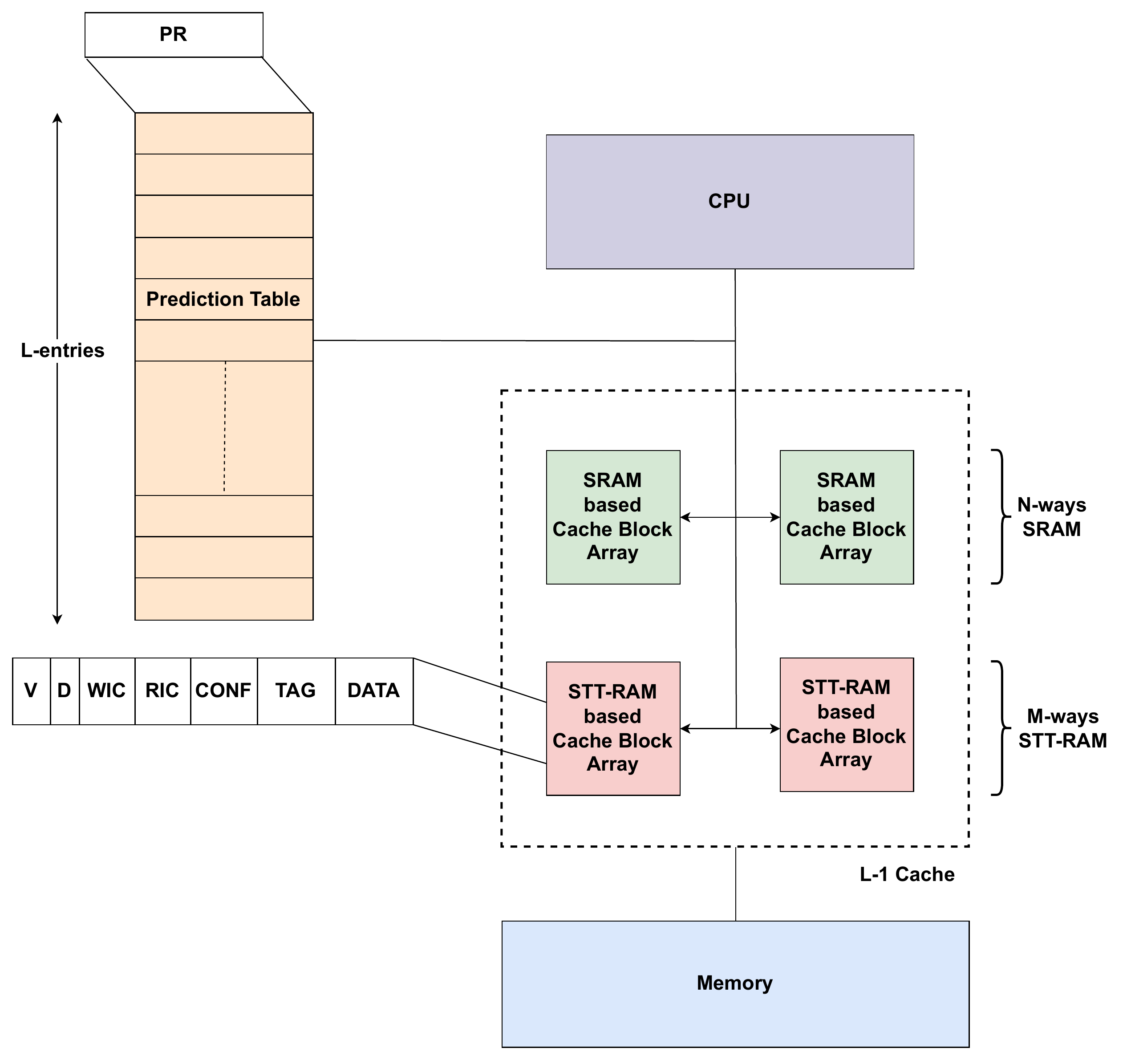}
\caption{Overview of the Proposed Architecture}
\label{Fig1}
\end{figure}

The proposed architecture is shown in figure \ref{Fig1}.

Every cache set in the proposed architecture contains a mix of SRAM and STT-RAM cache blocks. Along with the valid bit (V), dirty bit (D), tag, and data in each cache block, we added three more entries: i) Read-Intensive Counter (RIC), ii) Write-Intensive Counter (WIC), and iii) Confidence bits (CONF). These three entries are beneficial for cache placement and migration policies. We classified blocks into two types: read-intensive blocks and write-intensive blocks. Read-intensive (RI) blocks have more read accesses than a predefined threshold at a given point in time, while write-intensive (WI) blocks have more write accesses than a predefined threshold at a certain point in time.

We keep two counters for each block, RIC and the WIC. Furthermore, we added a 2-bit CONF field that tracks important blocks; important block information is helpful during power failures. A prediction table has also been included. Each prediction table entry has a previous region (PR) bit. During the replacement/eviction process, this PR bit is updated. The PR bit stores the block's most recent cache region.

\subsection{Placement and Migration Policies} \label{PMP}



We describe the proposed block placement and migration policies in this section. Because STT-RAM has higher read/write latency and consumes more energy than SRAM, the placement policy aims to reduce the number of writes to STT-RAM. STT-RAM write latency is ten times more than its read latency \cite{jog}. Therefore, we would like to place write-intensive blocks in the SRAM cache and read-intensive blocks in the STT-RAM cache. We use the PR bit to check the prediction table and place the block in the appropriate cache region based on whether it is read-intensive or write-intensive.

Algorithm \ref{alg3} demonstrates placement policy in case of a cache miss. Line 1 uses a tag to check the prediction table on a read/write miss. We access the PR bit associated with the tag entry. We keep the PR bit to note the previous block placement information for that tag entry. If PR=0, line 3-5 in algorithm \ref{alg3} checks whether the corresponding STT-RAM cache set is full or not. If it is full, we replace the block with the lowest RIC value; otherwise, we place it in the STT-RAM cache. Suppose PR !=0, line 10-12 in algorithm \ref{alg3} checks whether the SRAM cache set is full or not. We replace the block with the lowest WIC value; else, we place the block in the SRAM cache.

\begin{algorithm}
{\fontsize{10pt}{10pt}\selectfont
\caption{ Placement Algorithm in case of Cache miss }
\label{alg3}
\begin{algorithmic}[1]
\State Check Prediction Table.
\If{ $PR == 0$} 
    \If{STT-RAM set is full}
        \State Replace block with lowest $b.RIC$ 
        \State Update the replaced block's PR bit in the Prediction Table.
\Else
        \State Place in the STT-RAM cache.
        \State Re-Intialize b.RIC, b.WIC to zero.
\EndIf
\Else
        \If{SRAM set is full}
            \State Replace block with lowest b.WIC.
            \State Update the replaced block's PR bit in the Prediction Table.
        \Else
            \State Place in the SRAM cache.
            \State Re-Initialize b.RIC, b.WIC to zero.
        \EndIf
        \EndIf
\end{algorithmic}
}
\end{algorithm}


Algorithm \ref{alg1} describes placement and migration policies whenever there is a read hit. Line 1 checks the block's RIC value with the empirically determined threshold. We fixed the threshold limit empirically. If the block's RIC is equal to the threshold, we call that block an RI block. The proposed placement policy suggests that all RI blocks should place in STT-RAM. If the block is present in the SRAM cache, we migrate from SRAM to STT-RAM and re-initialize RIC, WIC, and CONF to zero. If the block is not in the SRAM cache, we place the block in the STT-RAM cache and increment CONF by 1. If the threshold does not equal the block's RIC value, we increment RIC by 1. The block chosen for replacement has to update its PR bit in the prediction table. If RIC reaches the threshold and CONF reaches 11 state, then we don't increment RIC.

\begin{algorithm}
{\fontsize{10pt}{10pt}\selectfont
\caption{ Placement and Migration Algorithm in case of Read hit  }
\label{alg1}
\begin{algorithmic}[1]
\If{ b.RIC == threshold } 
    \If{Block is in SRAM}
    \If{STT-RAM set is full}
        \State Replace block with lowest b.RIC.
        \State Update the replaced block's PR bit in the Prediction Table.
    \State Migrate to STT-RAM.
  \Else
    \State Migrate to STT-RAM.
    \EndIf
\State Re-Initialize b.RIC, b.WIC, b.CONF to zero.
 \EndIf
 \State b.CONF = b.CONF + 1
 \State Re-Initialize b.RIC to zero.
 \Else
  \State b.RIC = b.RIC + 1
\EndIf
\end{algorithmic}
}
\end{algorithm}


Algorithm \ref{alg2} describes placement and migration policies whenever there is a write hit. Line 1 checks the block's WIC value with the threshold. If the block's WIC equals the threshold, we call that block a WI block. The proposed placement policy suggests that all WI blocks should place in SRAM. If the block is already present in the STT-RAM cache, we migrate from STT-RAM to SRAM and re-initialize RIC, WIC, and CONF to zero. This case reduces the number of writes to the STT-RAM cache. If the threshold does not equal the block's WIC value, we increment WIC by 1. The block chosen for replacement has to update its PR bit in the prediction table. If WIC reaches the threshold and CONF reaches 11 state, then we don't increment WIC.

\begin{algorithm}[htb]
 {\fontsize{10pt}{10pt}\selectfont
 \caption{ Placement and Migration Algorithm in case of Write hit }
 \label{alg2}
 \begin{algorithmic}[1]
 \If{ b.WIC == threshold }
  \If{Block is in STT-RAM }
  \If{ SRAM set is full}
\State Replace block with lowest b.WIC.
  \State Update the replaced block's PR bit in the Prediction Table.
   \State Migrate to SRAM.
\Else
   \State Migrate to SRAM.
  \EndIf
  \State Re-Initialize b.RIC, b.WIC, b.CONF to zero.
  \EndIf
   \State b.CONF = b.CONF + 1
    \State Re-Initialize b.WIC to zero.
\Else
  \State b.WIC  = b.WIC + 1;
 \EndIf
\end{algorithmic} 
 }
 \end{algorithm}

 \subsection{Prediction Table Design} \label{PTD}


The importance of the prediction table in the proposed architecture is to store the previous region for the respective tag entry. The prediction table has L entries, where L denotes the number of entries in the prediction table. This table acts as a direct-mapped buffer, indexed using (Address/block\_size) \% L. Each entry in the prediction table has a PR (Previous Region) field. The prediction table does not store the tag bits in order to save area; its size is L bits. Initially, all bits in the prediction table are set to 1.

We update the PR field whenever there is a replacement in the cache due to the SRAM/STT-RAM set being full. If PR is 1, the block is a WIC because its WIC was greater than RIC during the replacement. Place the WIC block into the SRAM cache region. If PR is zero, the block is a RIC because its RIC is greater than WIC during the replacement. Place the RIC block into the STT-RAM cache region.

\subsection{Support for Intermittent Power Supply} \label{IPS}

Our proposed architecture supports intermittent computing and performs well during frequent power failures. We define important blocks as those with high CONF values. We use RIC/WIC values to update the CONF field. When power is restored in a traditional architecture, we begin execution by accessing blocks from the main memory and copying them to the cache. We save important blocks in STT-RAM that help to start execution without restoring blocks from the main memory to SRAM.

\begin{figure}[htb]
\includegraphics[width= 1\linewidth]{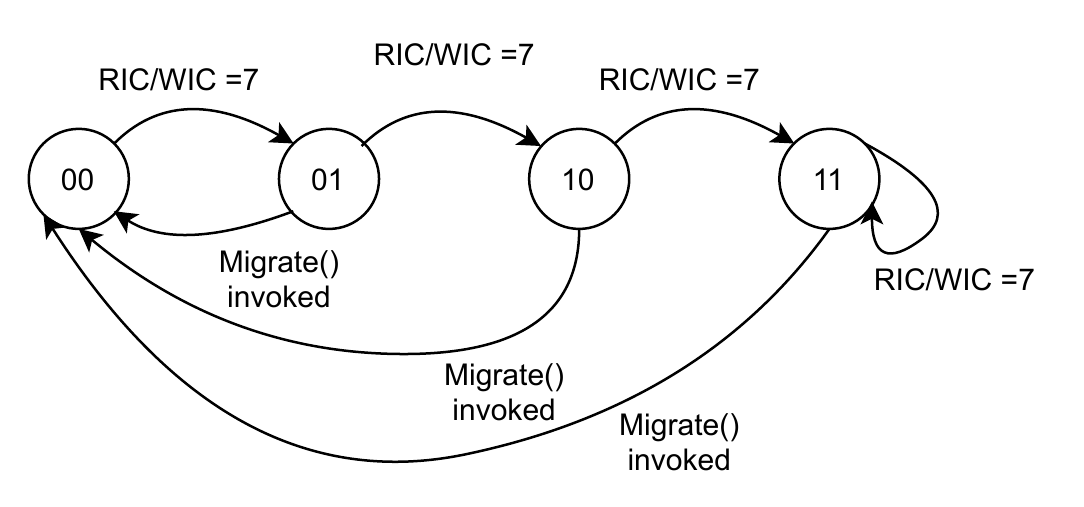}
\caption{State Diagram for Updating CONF}
\label{inter21}
\end{figure}

We propose a state model to assist in determining the most important blocks. Using the CONF field, we can determine which blocks should be present in STT-RAM during a power failure. Initially, CONF is in 00 state and supports four states, i.e., 00, 01, 10, and 11 states, as shown in figure \ref{inter21}. To represent the proposed state model, we need a 2-bit CONF field. The algorithmic process of updating the CONF field has already been described in algorithm \ref{alg1} and \ref{alg2}.

In summary, if RIC/WIC exceeds the threshold, CONF is increased by one and advances to the next state. When CONF is in the 11 state and crosses the threshold, it remains in that state. If any migration happens from SRAM/STT-RAM to STT-RAM/SRAM cache, then CONF resets to 00 state along with the RIC and WIC values.

During a power failure, the proposed backup policy triggers to save important blocks from SRAM to STT-RAM. According to the proposed backup policy, the blocks with the CONF field 11 are the most important blocks. Therefore, we prioritize blocks with the CONF field in the order of 11 $>$ 10 $>$ 01 $>$ 00. If any SRAM block has a CONF field of 11, we replace that block with the least priority block in STT-RAM. If there is no block with 11 state in the SRAM, we decrement our priority order by 1.

Now our priority becomes 10. If there are blocks with 10 state in the SRAM cache line, we replace the blocks with the least priority block in STT-RAM. If there is no block with 10 state in the SRAM line, we decrement our priority order by one. Similarly, we check blocks with 01 and 00 states. Priority with 00 is the case where we copy the SRAM contents to STT-RAM and STT-RAM contents to PCM. Whenever power comes back, STT-RAM contents are accessed automatically without copying to SRAM. Our migration policy automatically migrates from STT-RAM to SRAM if needed and vice-versa.

\subsection{Detailed Example} \label{DE}

\begin{figure}[htb]
\includegraphics[width= 1\linewidth]{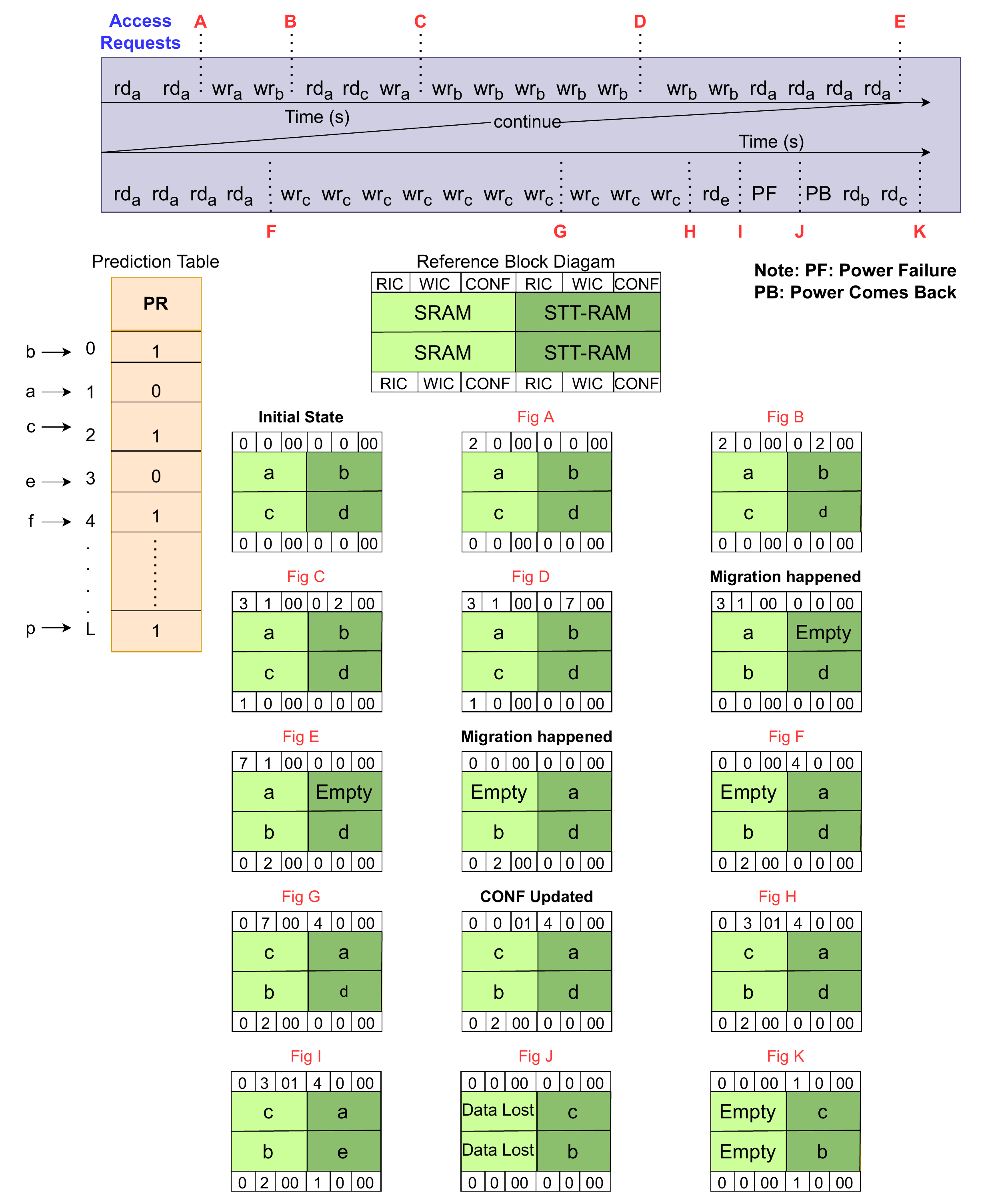}
\caption{Working Example of the Proposed Architecture}
\label{inter31}
\end{figure}

Figure \ref{inter31} illustrates the detailed working of the proposed architecture. In figure \ref{inter31}, Initially, the SRAM cache has (a,c) blocks, and the STT-RAM cache has (b,d) blocks. We defined all counters and CONF as a tuple [RIC, WC, CONF] and initialized it to [0, 0, 00]. A prediction table has a PR field. We take a sequence of access requests; read requests are labeled as $rd_i$ (i.e., read block i), and write requests are labeled as $wr_i$ (i.e., write block i). We labeled different timing points as A, B, C, .., K. In this section, we discuss how the proposed architecture works after every timing point.

In Fig A of figure \ref{inter31}, we update the RIC of `a' to 2 because of two consecutive reads. In Fig B, the WIC of `b' has become 2. In Fig C, [RIC, WIC] of `a' updates to [3, 1]. In Fig D, the WIC of `b' becomes 7, which equals the threshold and becomes a write-intensive block. Our placement policy suggests that write-intensive blocks should place in SRAM. SRAM set is full; to replace the block, we find the block having the lowest WIC. Block `c' has a low WIC value; we replace `c' with `b' and reset all `b' counters to [0, 0, 00]. In Fig E, the RIC of `a' becomes 7, which equals the threshold and becomes a read-intensive block. Our placement policy suggests that read-intensive blocks should place in STT-RAM. STT-RAM set had one empty slot; we migrated `a' from SRAM to STT-RAM. Reset all `a' counters to [0, 0, 00] and update the WIC of `b' to 2.

In Fig F, the RIC of `a' updates to 4. A new block request `c' occurred between F and G timing points. Block request `c' is not present in both caches. Check the prediction table for index 2, associated with $tag_c$, to find the c's PR field. We found an entry in the prediction table of index 2 with PR = 1. If the PR value is 1, the block is placed in the SRAM cache during the last eviction. We place `c' in the SRAM cache. In Fig G, the WIC of `c' updates to 7, which equals the threshold, becomes a write-intensive block, and updates the RIC of `a' to 4. Our placement policy suggests that write-intensive blocks should be placed in the SRAM; `c' is already in SRAM. We update the CONF of `c' to 01 and reset the counter values. After H, WIC of `c' updates to 3. 

A new block request `e' occurred between H and I timing points. Block request `e' is not present in both caches. Check the prediction table for index 3, associated with $tag_e$, to find the e's PR field. We found an entry in the prediction table of index 3 with PR = 0. If the PR value is 0, the block is placed in the STT-RAM cache during the last eviction. We place `e' in the STT-RAM cache. STT-RAM set is full; we find the lowest RIC to replace the block. Block `d' has a low RIC value; we replace `e' with `d'. Update all `e' counters to \{1, 0, 00\}. 

Power failure (PF) occurred; our backup policy saves important blocks using the CONF field. Where the CONF of  `c' has 01 and `a' has 00, our priority order suggests that 01 has the highest priority than 00. We place `a' to the main memory and backup `c' to STT-RAM. We prefer write-intensive blocks compared to read-intensive during a power failure. So `b' replaces `e'. In Fig J, `c' and `b' are saved to STT-RAM. Whenever power comes back (PB), we don't require any restoration process. Fig K shows the RIC of `c' and `b' updates to 1.

\subsection{Storage Overhead} \label{SO}

We analyze the storage overhead because we added extra bits, a prediction table, and backup logic. For the same system configuration shown in table \ref{tab1}, we evaluate the area overhead for the proposed architecture. We showed the area overhead as an example. There are two aspects of the proposed architecture that cause storage overhead.

 \begin{itemize}
     \item The proposed architecture has two 3-bit counters and two confidence bits per block. The data cache has 256 blocks, each with 8 bits, so the data cache requires 256*8=2048 bits.
     
     \item The proposed prediction table has 4K byte entries with 1-bit per entry, resulting in a total storage overhead of 1024*4 = 4096 bits.

 \end{itemize}
 
The overall storage overhead of the proposed architecture will be 2048 + 4096 = 6144 bits=0.75KB. The total percentage of area overhead is about 0.75KB/32KB=2.34\%.

\section{Experimental Setup and Results} \label{p5}
\subsection{Experimental Setup}
We evaluate the proposed architecture using the gem5 \cite{13} simulator and 18 benchmarks from the MiBench suite \cite{mibench}. Overall micro-architectural parameters used for implementation are shown in table \ref{tab1}.

\begin{table}[htb]
\centering
\caption{System Configuration}
\label{tab1}
\begin{tabular}{|l|l|}
\hline
\textbf{Component}   & \textbf{Description}             \\ \hline
\textbf{CPU core}    & 1-core, 480MHZ               \\ \hline
\textbf{L1 Cache} &
  \begin{tabular}[c]{@{}l@{}}Block size - 64-byte, 4-way associative (2-way SRAM, \\ 2-way STT-RAM);\\ Private cache (16KB hybrid D-cache, and 16KB I-cache)\end{tabular} \\ \hline
\textbf{Size Parameters} &
  \begin{tabular}[c]{@{}l@{}}VB-1bit, WIC and RIC-3bits, CONF-2bits, \\ L- 4K bytes, threshold-7, and PR-1bit\end{tabular} \\ \hline
\textbf{Main memory} & 128MB PCRAM \\ \hline
\textbf{Others} & \begin{tabular}[c]{@{}l@{}} Clock Period: 2ns, \\SRAM Read: 1 Cycle, \\ SRAM Write: 2 Cycles,\\ STT-RAM Read: 2 Cycles, \\ STT-RAM Write: 10 Cycles, \\ PCM Read: 35 Cycles, and \\ PCM Write: 100 Cycles \end{tabular} \\ \hline
\end{tabular}
\end{table}

\begin{table}[htb]
\centering
\caption{Nvsim parameters of SRAM, MRAM Caches, and PCM memory (350K, 22nm)}
\label{tab2}
\begin{tabular}{|l|l|l|l|l|}
\hline
\textbf{Parameter} & \textbf{\begin{tabular}[c]{@{}l@{}}16KB \\ SRAM\end{tabular}} & \textbf{\begin{tabular}[c]{@{}l@{}}16KB \\ MRAM\end{tabular}} & \textbf{\begin{tabular}[c]{@{}l@{}}128MB \\ PCRAM\end{tabular}} \\ \hline
\textbf{Read Latency} & 0.792 ns & 1.994 ns & 204.584 ns \\ \hline
\textbf{Read Energy} & 0.006 nJ & 0.081 nJ & 1.553 nJ \\ \hline
\multirow{2}{*}{\textbf{Write Latency}} & \multirow{2}{*}{0.772 ns} & \multirow{2}{*}{10.520 ns} & RESET - 134.954 ns \\ &  &  & SET - 264.954 ns \\ \hline
\multirow{2}{*}{\textbf{Write Energy}} & \multirow{2}{*}{0.002 nJ} & \multirow{2}{*}{0.217 nJ} & RESET - 6.946 nJ \\ 
 &  &  & SET - 6.927 nJ \\ \hline
\textbf{Leakage Power } & 18.972 mW  & 3.014 mW & - \\ \hline
\end{tabular}
\end{table}




Table \ref{tab2} shows the dynamic energy and latency for a single read and write operation to SRAM and STT-RAM, taken using Nvsim \cite{25}.

\subsection{Baseline Architecture} \label{p51}
We modeled a baseline architecture to compare with the proposed architecture. 

We first compared the performance and dynamic energy consumption of pure SRAM, pure STT-RAM, and hybrid (SRAM and STT-RAM) cache architectures to determine the baseline architecture. Based on the analysis, we choose the relevant baseline architecture to compare the proposed and existing architectures throughout this work.

\begin{figure}[htb]
    \centering
    \subfloat[\centering Execution Time]{{\includegraphics[width=1\linewidth]{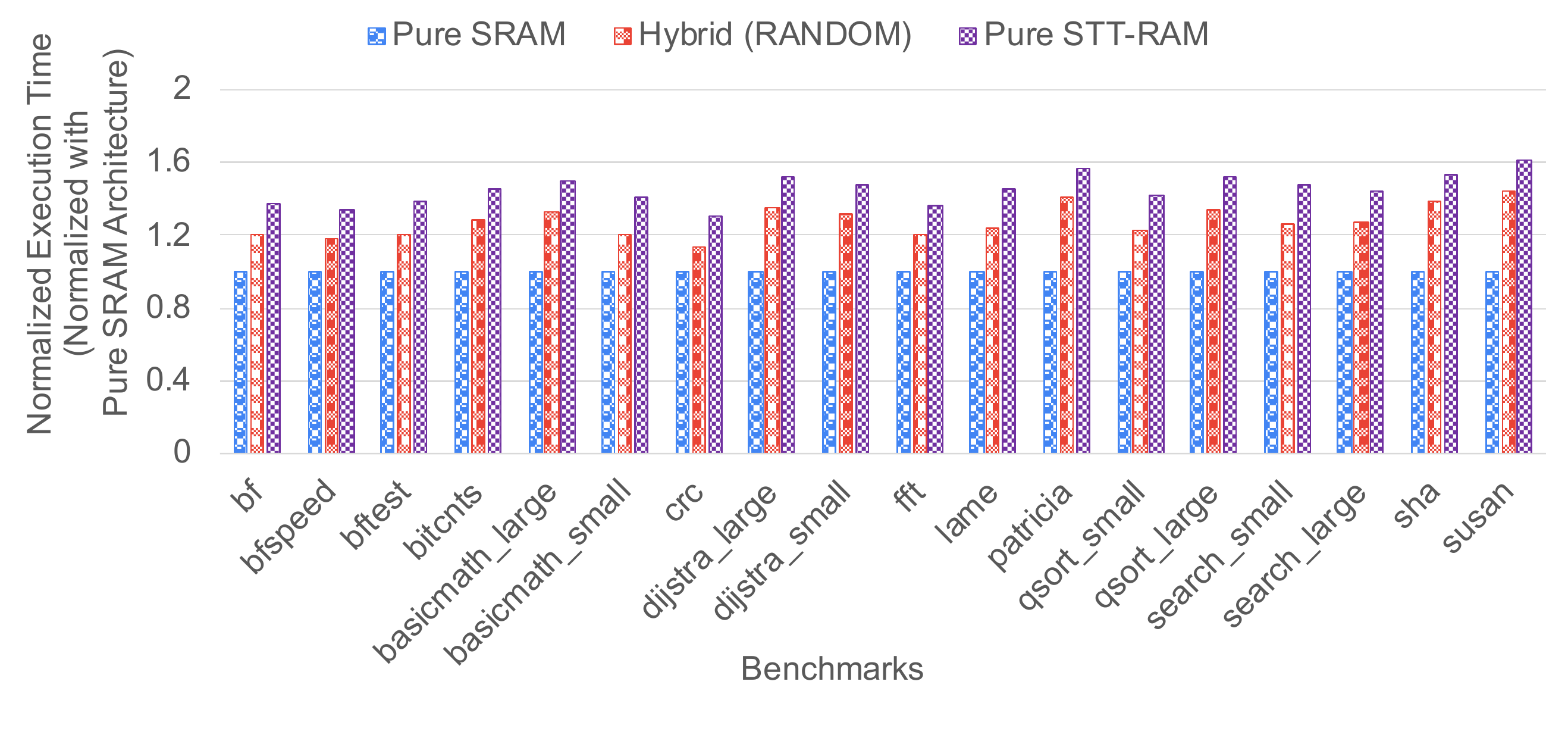} }}%
    \qquad
    \subfloat[\centering Dynamic Energy Consumption]{{\includegraphics[width=1\linewidth]{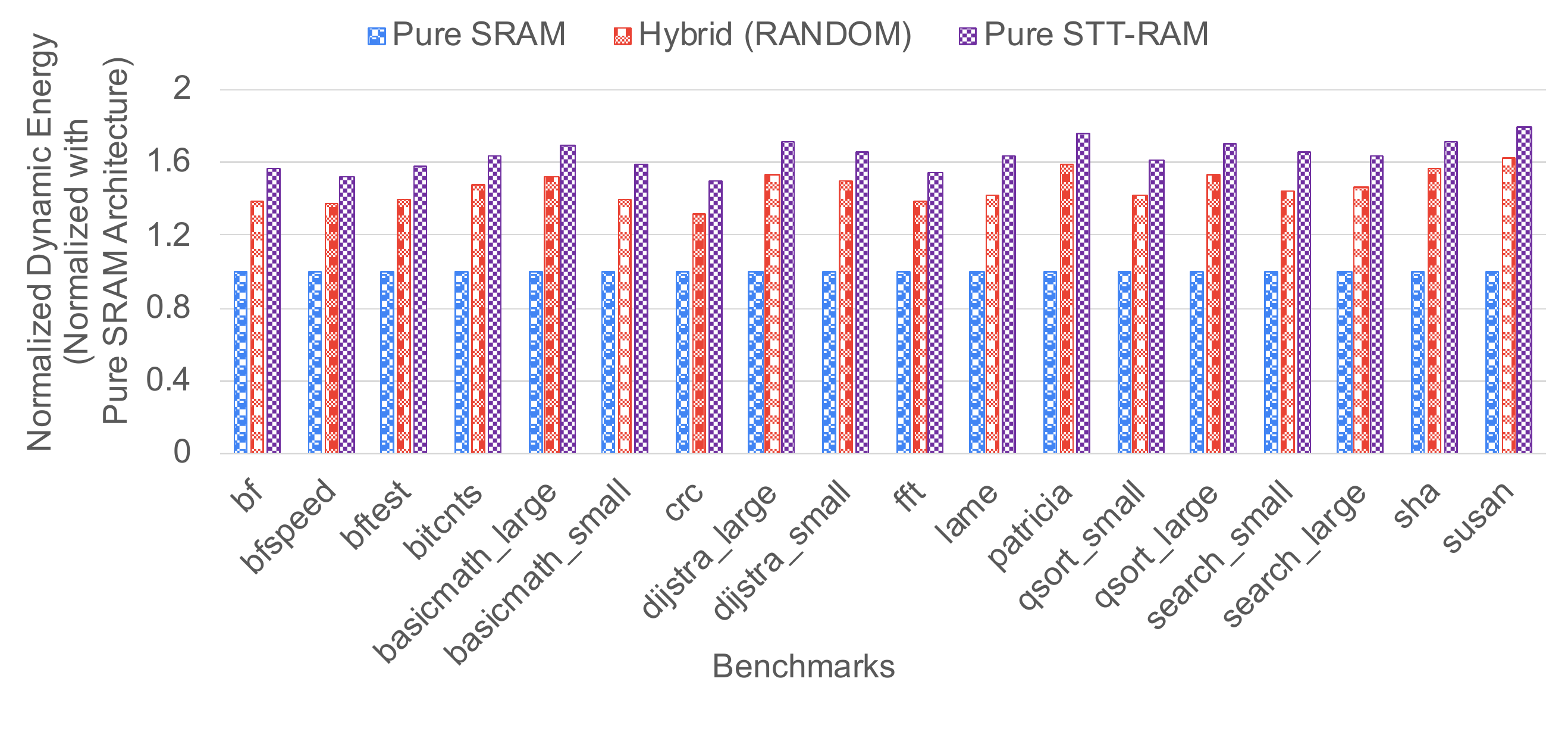} }}%
    \caption{Comparisons between Pure SRAM, Pure STT-RAM, and Hybrid Cache Architectures.}%
    \label{Fig2}%
\end{figure}

\begin{itemize}
     \item \textbf{Pure SRAM cache}: We don't require any placement or migration policies in pure SRAM cache because we have only SRAM at L1.

    \item \textbf{Pure STT-RAM cache}: We don't require any placement or migration policies in pure STT-RAM cache because we have only STT-RAM at L1.
   \item \textbf{Hybrid cache}: At L1, the hybrid cache architecture includes both SRAM and STT-RAM. We use a random placement policy in this HCA. In the random placement policy, the block is randomly placed in either SRAM or STT-RAM. We use the migration policy that moves blocks from one cache to another based on counters. We empirically determined the threshold as 7 and the size of the counters as 3 bits. Assume the WIC exceeds the threshold and is present in the STT-RAM cache region. In that case, we migrate that block into the SRAM cache region. Assume the RIC exceeds the threshold value and is found in the SRAM cache region. In that case, we migrate that block into the STT-RAM cache region. 
   
\end{itemize}


We set the L1 size to 32KB in all three architectures. Above all three cache architectures, we did not use any prediction mechanisms. In this figure \ref{Fig2}, the performance and energy consumption of the cache architectures are normalized with the pure SRAM cache architecture. As shown in figure \ref{Fig2}, hybrid-based architecture performs in between pure SRAM and pure STT-RAM cache architectures, i.e., hybrid-based architecture is better than pure STT-RAM cache. 

\textbf{Baseline Architecture:} We selected hybrid-based architecture as our baseline architecture throughout this paper with the above-mentioned modeling details.

\begin{figure}[htb]
  \includegraphics[width= 1\linewidth]{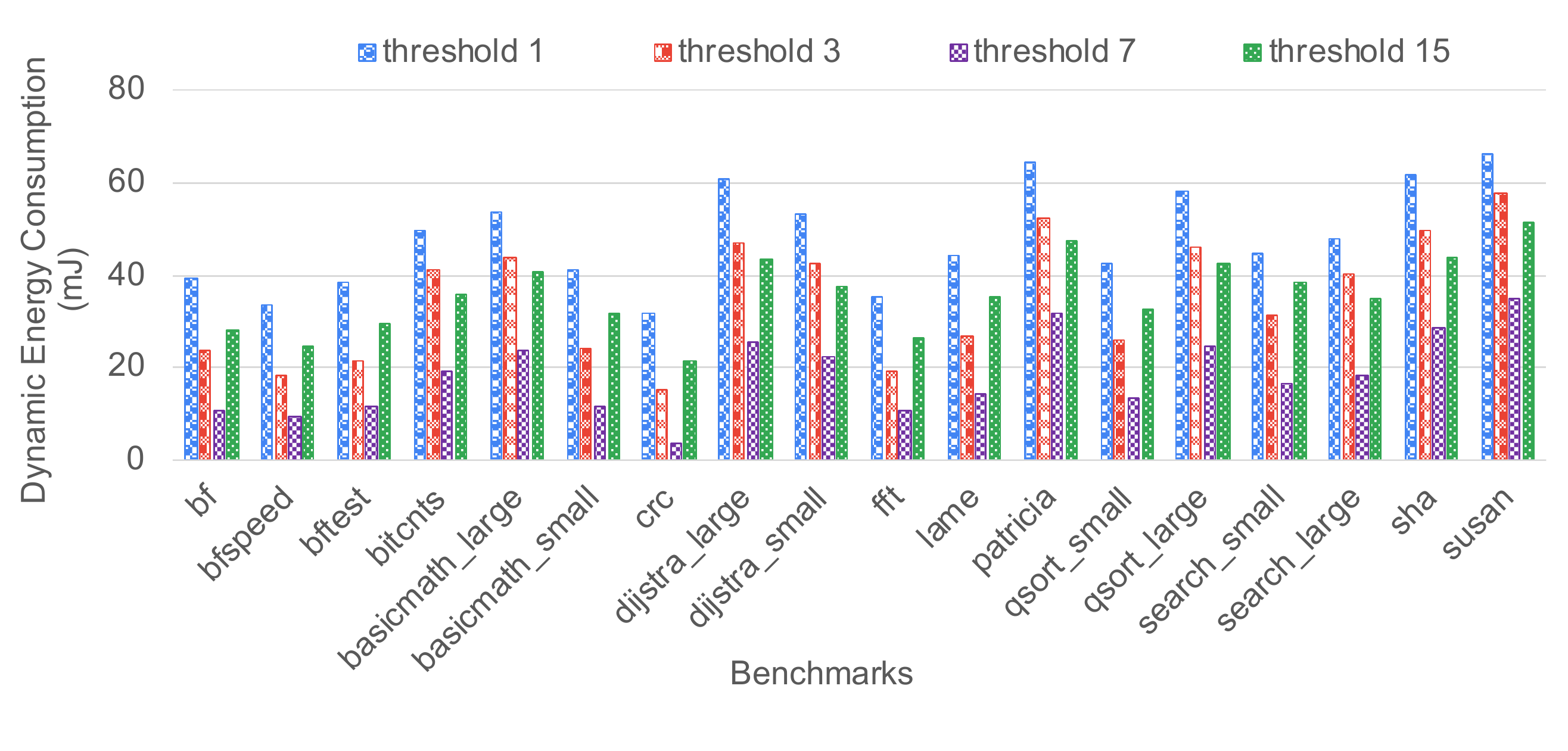}
\caption{Dynamic Energy consumption for Various Threshold Values }
\label{tab32}
\end{figure}

We experimented with the baseline architecture to determine the threshold value. When the respective counter crosses its threshold, we move the block from one cache to another to check energy values. The size of the counters is determined by the threshold value. For example, if the threshold is 3, the counter size is $log4$. (counts from 0 to 3). We experimented with threshold values of 1, 3, 7, and 15. We observed that threshold value 7 consumes less energy than the other threshold values on average. We set the threshold to 7 because we noticed that migrations between cache regions increase when the threshold is exceeded. We also observed that NVM gets more writes if the threshold is higher than 7, increasing HCA's energy consumption. Figure \ref{tab32} shows that threshold value 7 consumed less energy than the other threshold values. The threshold value in our proposed architecture is set to 7 throughout this work for the system configuration shown in table \ref{tab32}. We also performed experiments to analyze the selected threshold behavior on our proposed architecture in section \ref{sp}.

\subsection{ Results}

This section evaluates the proposed architecture under stable power and during intermittent power supply. We also evaluate the proposed architecture efficiency w.r.t traditional checkpointing approach under stable power and frequent power failures. Lastly, we evaluate the proposed architecture for $\eta$, $\theta$ w.r.t baseline, and existing architectures.

\subsubsection{\textbf{Under Stable Power}} \label{sp}

\begin{figure}[htb]
  \includegraphics[width= 1\linewidth]{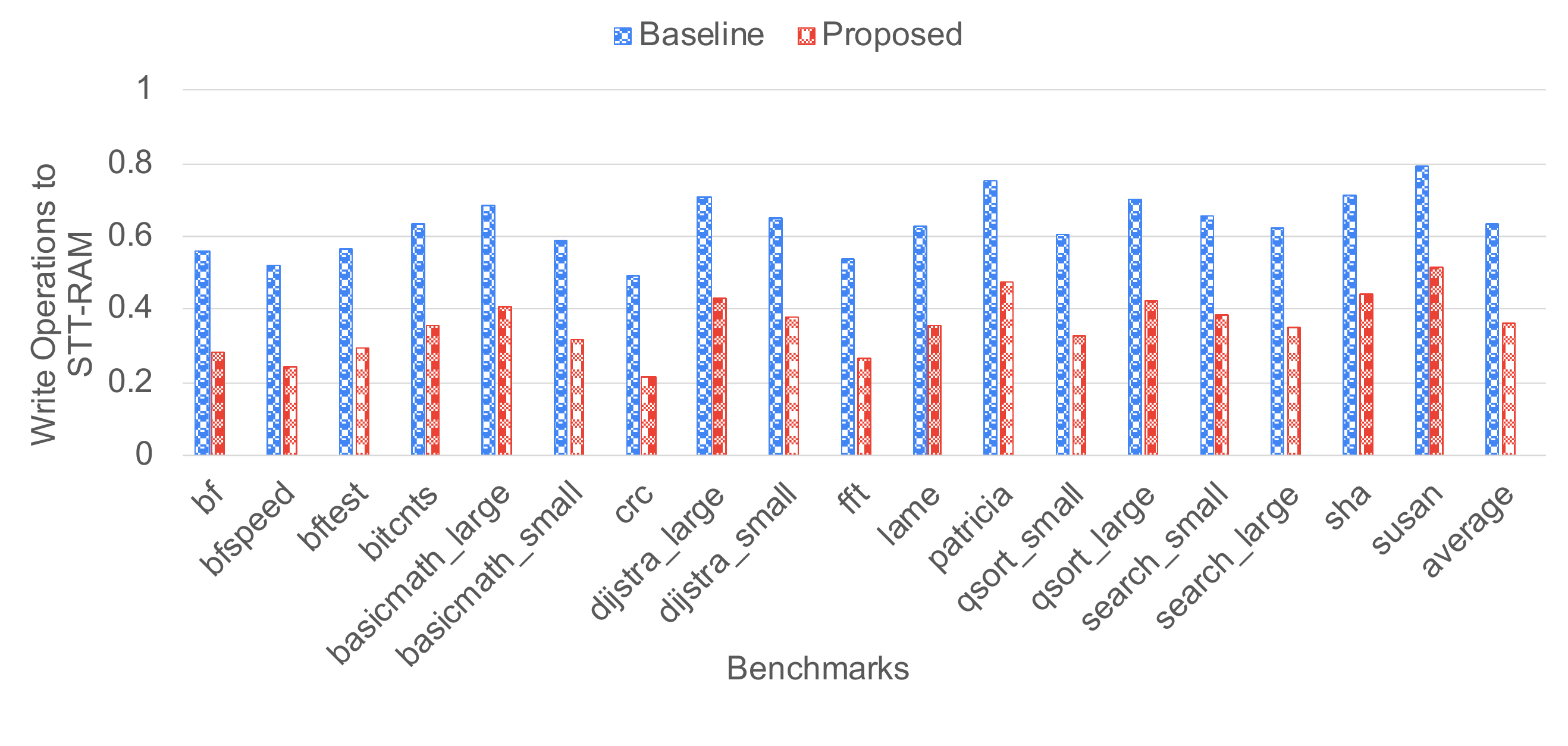}
\caption{ Write operations to \small{STT-RAM} }
\label{tab3}
\end{figure}

We compare our proposed architecture with the baseline and the architecture proposed by Xie et al. \cite{12}. We implemented Xie et al. \cite{12} work to analyze both stable power and intermittent power systems. For a fair comparison, all these architectures use the same system configuration shown in table  \ref{tab1} and energy/delay values of STT-RAM and PCM from table \ref{tab2}.

One of the main objectives of the proposed architecture is to reduce the number of writes to the STT-RAM cache. To achieve this, we place the write-intensive blocks in the SRAM cache. We have shown the ratio of the write operations to STT-RAM with total write accesses in figure \ref{tab3}. A lower number of writes to STT-RAM shows the effectiveness of the proposed architecture. The percentage of writes to the STT-RAM cache is normalized with the baseline architecture shown in figure \ref{tab3}. Overall, the proposed architecture helps in reducing STT-RAM write operations from 63.35\% to 35.93\% compared to the baseline architecture. 

\begin{figure}[htb]
  \includegraphics[width= 1\linewidth]{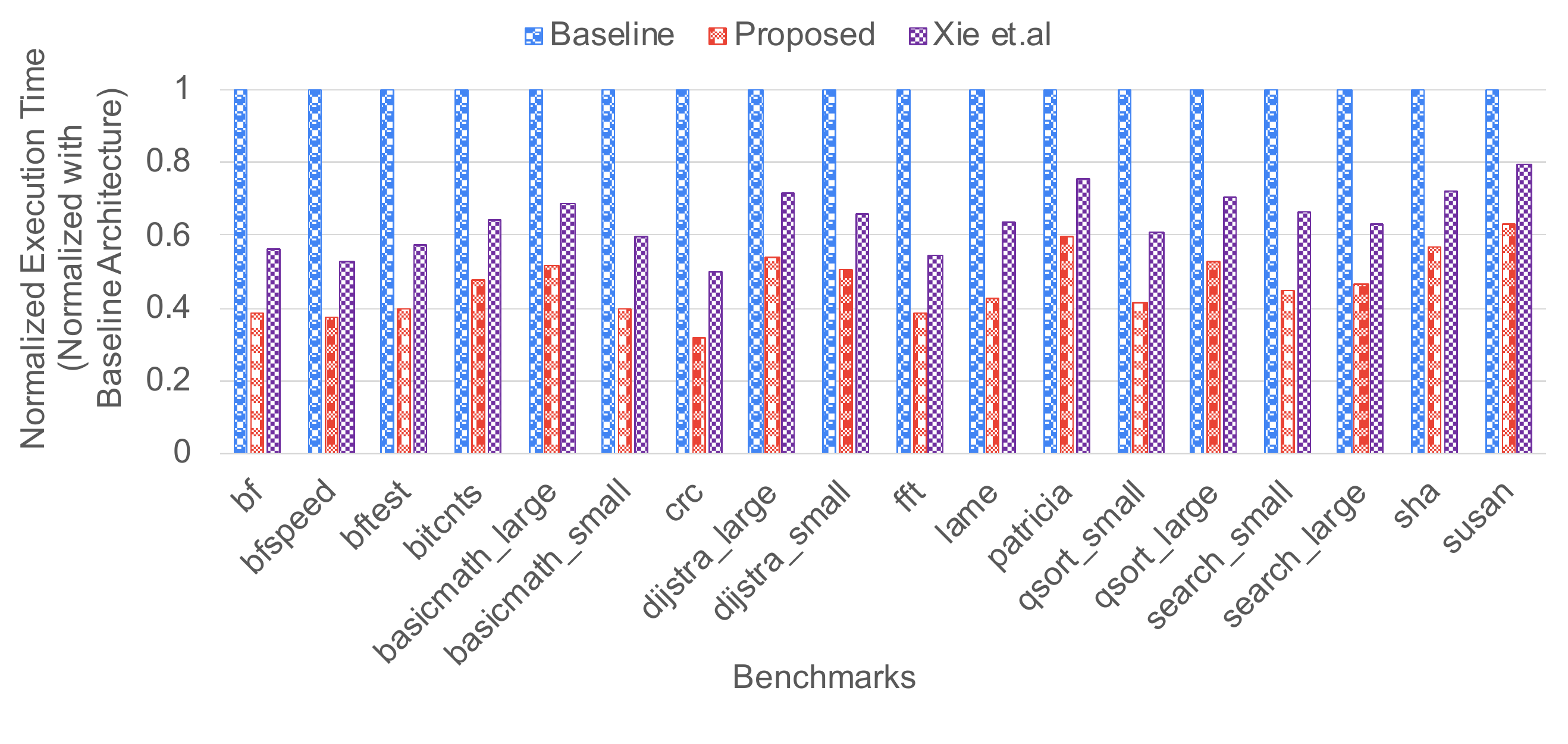}
\caption{ Comparisons between Proposed, Baseline, and Existing Architectures for Execution Time under Stable Power. }
\label{tab31a}
\end{figure}

\begin{figure}[htb]
  \includegraphics[width= 1\linewidth]{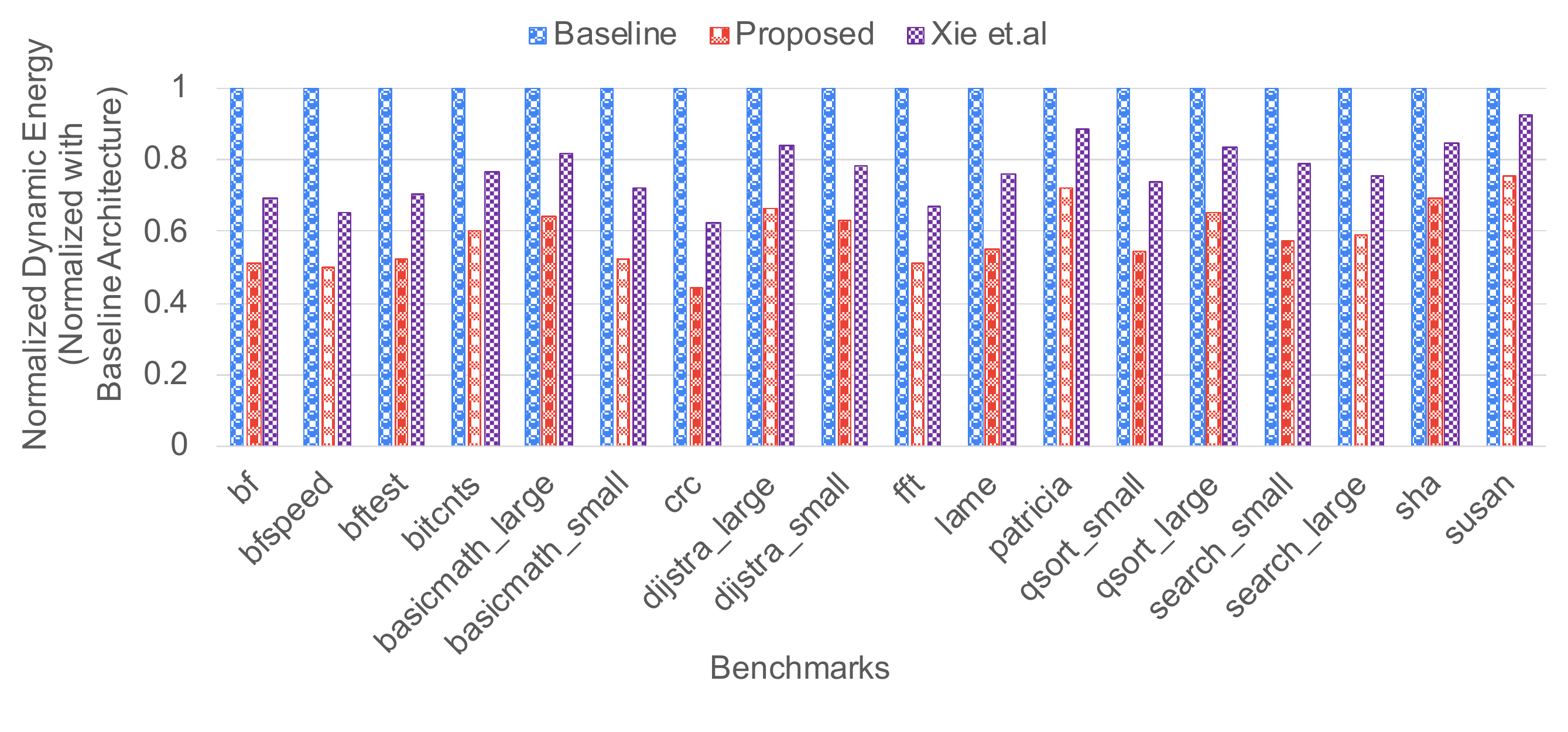}
\caption{ Comparisons between Proposed, Baseline, and Existing Architectures for Energy Consumption under Stable Power. }
\label{tab31b}
\end{figure}

Reducing the STT-RAM writes also guarantees better endurance and a lifetime of IoT nodes. The performance and energy consumption values are normalized with the baseline architecture in the figures, \ref{tab31a} and \ref{tab31b}. Figures \ref{tab31a} and figure \ref{tab31b} show better execution time and dynamic energy consumption than the baseline and existing architectures. We achieve better values because of accurate prediction when we compare the proposed architecture with Xie et al. architecture. Xie et al. \cite{12} work use a pattern sampler for prediction, but we maintained PR bit for every block in our proposed architecture. PR bit helps us with efficient block placement. If our prediction accuracy increases, the number of migrations decreases. If the number of migrations decreases, we observe fewer writes to NVM. Thus, our proposed architecture decreases the number of writes to STT-RAM.

\begin{figure}[htb]
  \includegraphics[width= 1\linewidth]{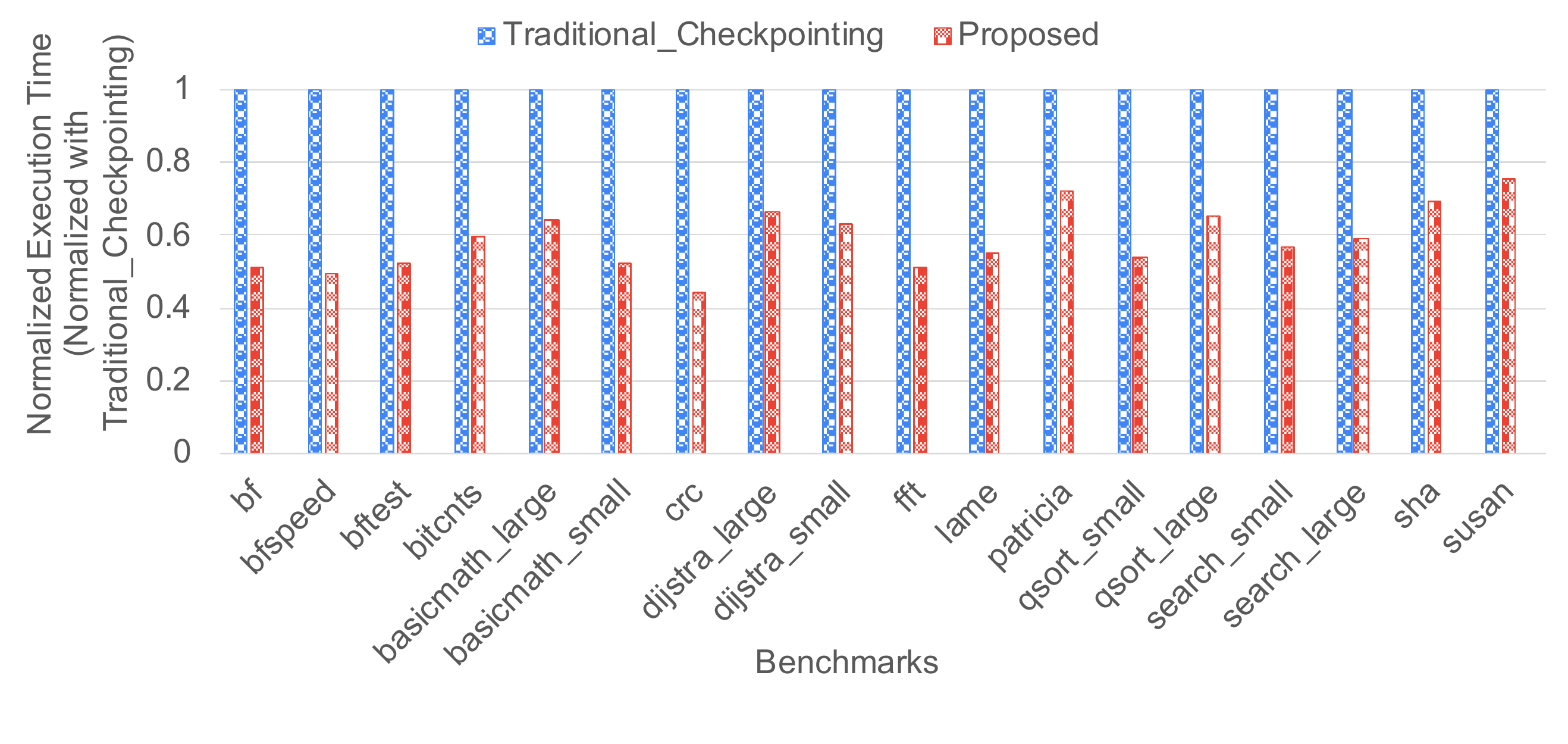}
\caption{ Comparison in terms of Performance Overhead during Stable Power}
\label{tab32a}
\end{figure}

Further, the proposed prediction table helps to decrease the number of migrations and accesses. Therefore, our architecture results in 32.85\% better execution time and saves 23.42\% of dynamic energy consumption than baseline architecture.

\begin{figure}[htb]
  \includegraphics[width= 1\linewidth]{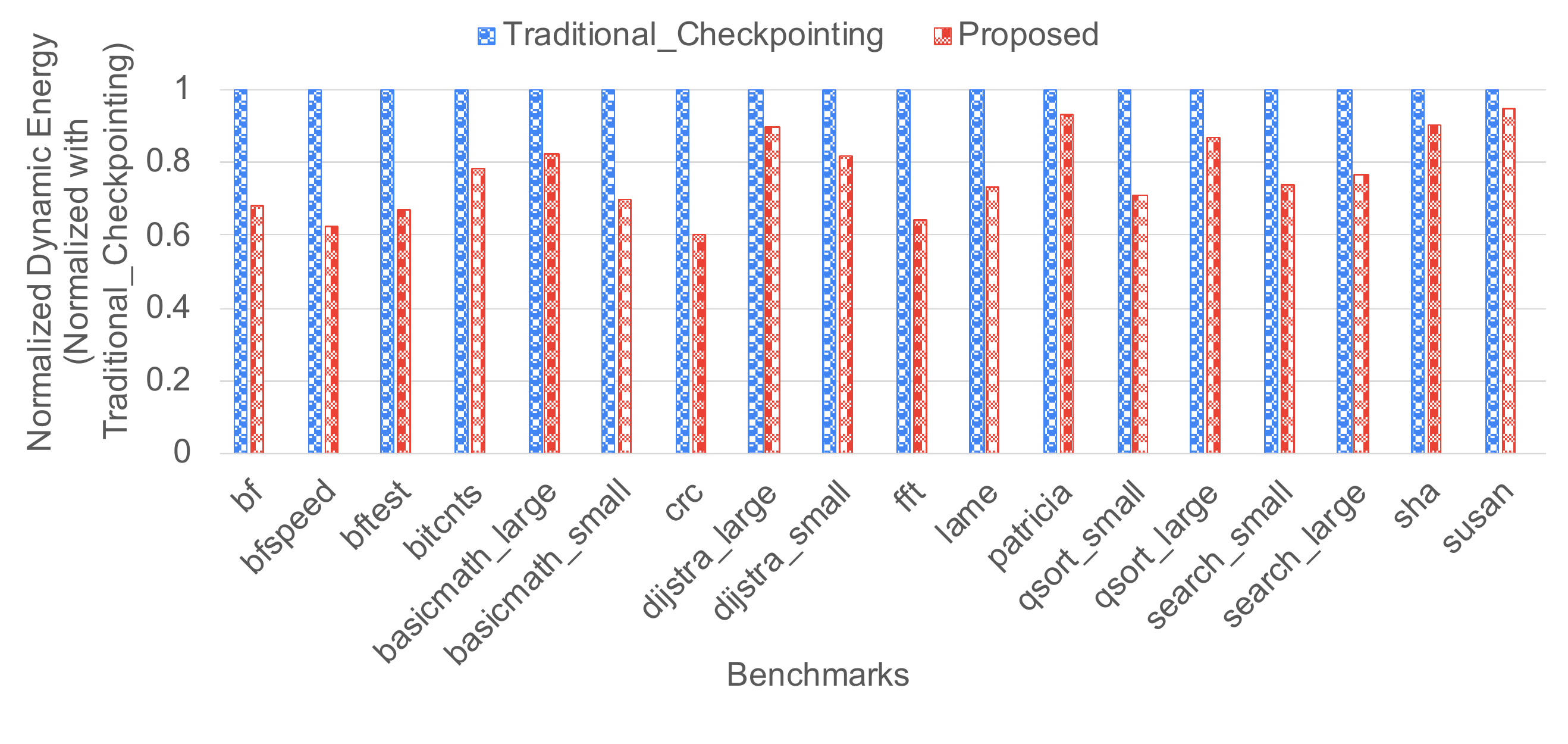}
\caption{ Comparison in terms of Energy Consumption during Stable Power }
\label{tab32b}
\end{figure}

During stable power, we performed experiments to compare the traditional checkpointing approach with the proposed architecture. We used a traditional checkpointing method, creating a safe point every 4 million instructions. We save the program state for every 4 million instructions to the main memory. Our proposed architecture outperforms traditional checkpointing. In traditional checkpointing, backup occurs for each safe point, but in the proposed architecture, backup occurs only during a power failure. We normalized the performance and energy consumption values with the traditional checkpointing approach. Proposed HCA reduces performance overhead and energy consumption by 21.03\% and 22.95\%, as shown in figures \ref{tab32a} and \ref{tab32b}.

\textbf{Analysis for the threshold value:} We performed experiments to analyze the selected threshold value with our proposed architecture in detail. For these experiments, we used five MiBench benchmarks with more instructions and writes, as shown in table \ref{tab411}, to better understand the relationship between different thresholds and dynamic energy consumption.

\begin{table}[htp]
\centering
\caption{Application Memory Patterns}
\label{tab411}
{
\begin{tabular}{|l|l|l|l|l|l|}
\hline
\textbf{Benchmarks} &
  \textbf{\begin{tabular}[c]{@{}l@{}} \#  of \\ Instructions \end{tabular}} &
  \textbf{\begin{tabular}[c]{@{}l@{}} \# of Mem \\ Reads \end{tabular}} &
  \textbf{\begin{tabular}[c]{@{}l@{}} \# of Mem \\ Writes \end{tabular}} &
  \textbf{Threshold} &
  \textbf{\begin{tabular}[c]{@{}l@{}}Dynamic \\ Energy \\ (mJ)\end{tabular}} \\ \hline\multirow{4}{*}{\textbf{qsort}}     & \multirow{4}{*}{469467835} & \multirow{4}{*}{73791088} & \multirow{4}{*}{60602059} & 1  & 89.53 \\  
                                           &                            &                           &                           & 3  & 67.19 \\  
                                           &                            &                           &                           & 7  & 68.98 \\  
                                           &                            &                           &                           & 15 & 76.11 \\ \hline
\multirow{4}{*}{\textbf{sha}}         & \multirow{4}{*}{49870857}  & \multirow{4}{*}{13422737} & \multirow{4}{*}{5004930}  & 1  & 103.15 \\  
                                           &                            &                           &                           & 3  & 92.33 \\  
                                           &                            &                           &                           & 7  & 71.60 \\  
                                           &                            &                           &                           & 15 & 89.79 \\ \hline

\multirow{4}{*}{\textbf{susan}}            & \multirow{4}{*}{111653876} & \multirow{4}{*}{24586223} & \multirow{4}{*}{10127992} & 1  & 116.09 \\  
                                           &                            &                           &                           & 3  & 109.67 \\  
                                           &                            &                           &                           & 7  & 91.45 \\  
                                           &                            &                           &                           & 15 & 89.21 \\ \hline
\multirow{4}{*}{\textbf{dijkstra}}          & \multirow{4}{*}{301988532} & \multirow{4}{*}{79999803} & \multirow{4}{*}{1045712}  & 1  & 102.19 \\  
                                           &                            &                           &                           & 3  & 98.77 \\  
                                           &                            &                           &                           & 7  & 84.10 \\  
                                           &                            &                           &                           & 15 & 96.07 \\ \hline
                                           
\multirow{4}{*}{\textbf{basicmath}} & \multirow{4}{*}{277951743} & \multirow{4}{*}{24217872} & \multirow{4}{*}{23164606} & 1  & 73.81 \\  
                                           &                            &                           &                           & 3  & 44.01 \\  
                                           &                            &                           &                           & 7  & 23.68 \\  
                                           &                            &                           &                           & 15 & 40.77 \\ \hline
\end{tabular}%
}
\end{table}

\begin{figure}[htb]
  \includegraphics[width= 1\linewidth]{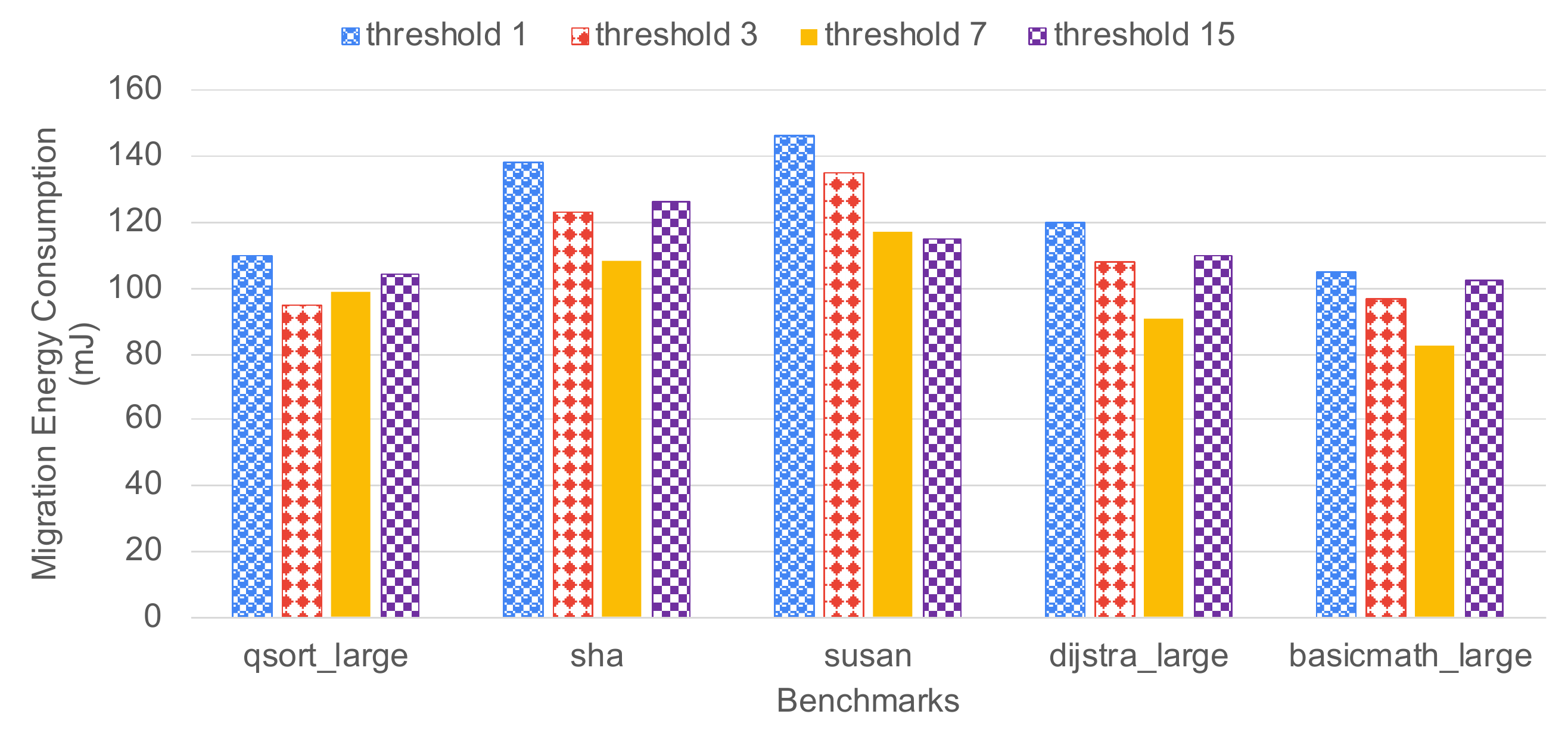}
\caption{ Migration Energy consumption for Various Threshold Values }
\label{Fig320}
\end{figure}

We observed that the number of migrations decreases when the threshold is at 7 for three out of five benchmarks. Figure \ref{Fig320} shows these five benchmarks' migration energy (the total energy required to migrate the blocks from SRAM/STT-RAM to SRAM/SRAM). The difference between threshold seven and other benchmarks was minimal for the remaining two benchmarks. The number of incorrect placements increases when we increment the counter size by 1 bit for many benchmarks, which increases the number of migrations between cache regions. As a result, for the considered system configuration, threshold 7 is beneficial for many benchmarks. The threshold value depends on the system configuration.

\textbf{Analysis for Different Cache Settings:} We also performed experiments by changing the system configurations like cache sizes and associativity that are different from the system configuration shown in table \ref{tab1}. We used 6 different cache settings for these experiments, as shown in table \ref{tab7}. 

\begin{table}[htp]
\centering
\caption{Different Cache Configurations used to Analyze Proposed Policies}
\label{tab7}
{
\begin{tabular}{|l|l|l|l|}
\hline
\textbf{Configuration} & \textbf{Cache Setting} & \textbf{Cache Size} & \textbf{Associativity}                                                      \\ \hline
1 & 16K (0:8)              & 16KB                & \begin{tabular}[c]{@{}l@{}}8-way\\ (0-way SRAM, 8-way STT-RAM)\end{tabular} \\ \hline

2 & 16K (2:6)              & 16KB                & \begin{tabular}[c]{@{}l@{}}8-way\\ (2-way SRAM, 6-way STT-RAM)\end{tabular} \\ \hline
3 & 16K (4:4)              & 16KB                & \begin{tabular}[c]{@{}l@{}}8-way\\ (4-way SRAM, 4-way STT-RAM)\end{tabular} \\ \hline
4 & 16K (6:2)              & 16KB                & \begin{tabular}[c]{@{}l@{}}8-way\\ (6-way SRAM, 2-way STT-RAM)\end{tabular} \\ \hline
5 & 16K (8:0)              & 16KB                & \begin{tabular}[c]{@{}l@{}}8-way\\ (8-way SRAM, 0-way STT-RAM)\end{tabular} \\ \hline
6 & 32K (0:8)              & 32KB                & \begin{tabular}[c]{@{}l@{}}8-way\\ (0-way SRAM, 8-way STT-RAM)\end{tabular} \\ \hline
7 & 32K (2:6)              & 32KB                & \begin{tabular}[c]{@{}l@{}}8-way\\ (2-way SRAM, 6-way STT-RAM)\end{tabular} \\ \hline
8 & 32K (4:4)              & 32KB                & \begin{tabular}[c]{@{}l@{}}8-way\\ (4-way SRAM, 4-way STT-RAM)\end{tabular} \\ \hline
9 & 32K (6:2)              & 32KB                & \begin{tabular}[c]{@{}l@{}}8-way\\ (6-way SRAM, 2-way STT-RAM)\end{tabular} \\ \hline
10 & 32K (8:0)              & 32KB                & \begin{tabular}[c]{@{}l@{}}8-way\\ (8-way SRAM, 0-way STT-RAM)\end{tabular} \\ \hline

\end{tabular}%
}
\end{table}

\begin{figure}[htb]
  \includegraphics[width= 1\linewidth]{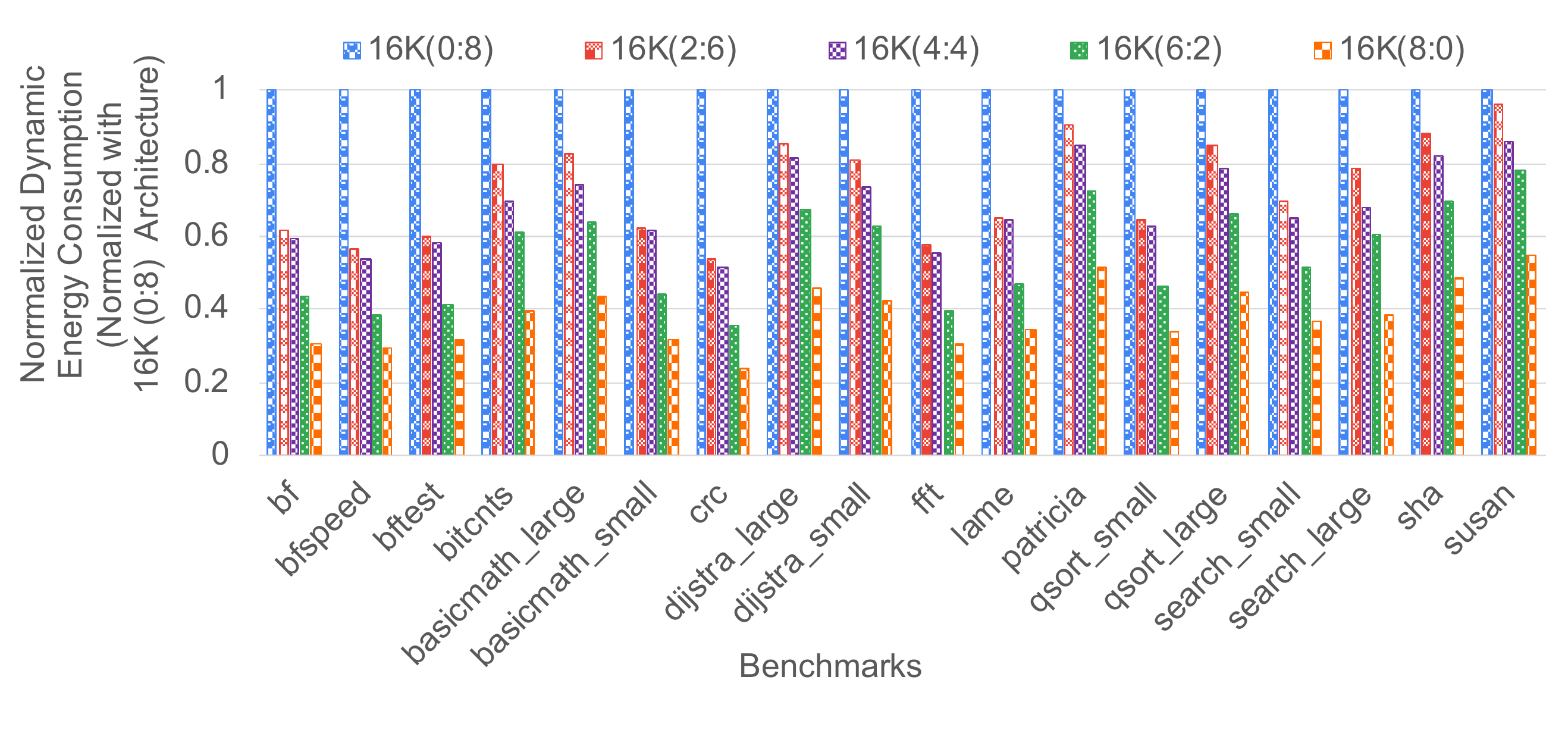}
\caption{ Dynamic Energy Consumption for Different Cache Configurations under Stable Power, where Cache Size is 16KB}
\label{Fig421}
\end{figure}

Here, we used two different cache sizes; one is 16 KB, and the other is 32 KB. We compared the energy consumption for these 10 configurations under stable power. We used all proposed policies and techniques in these 10 sets of configurations. In figure \ref{Fig421}, the energy consumption of the configurations \{2-5\} is normalized based on configuration-1 (pure STT-RAM-based cache architecture for 16KB cache size). 


We observed that for the 16KB cache size, configuration-1 consumes more energy than the other 4 configurations during stable power because STT-RAM consumes more energy than SRAM. During stable power supply, we observed that the configuration with more SRAM ways consumes less energy than others without relating to cache sizes. We observed 16K(8:0) setting consumes less energy than all other 4 configurations because it is like a pure SRAM-based architecture) and after this, the 16K(6:2) setting consumes less energy than all the other 3 configurations. After 16K(6:2), the 16K(4:4) setting consumes less energy than the other two configurations, i.e., 16K(2:6) and 16K(0:8). Compared to pure STT-RAM cache architecture, the 16K(6:2) setting consumes 38.10\% less energy, the 16K(4:4) setting consumes 17.97\% less energy, as shown in figure \ref{Fig421}. Compared to pure STT-RAM cache architecture, the 16K(2:6) setting consumes 5.82\% less energy, as shown in figure \ref{Fig421}.

Similarly, we observed 32K(8:0) setting consumes less energy than all other 4 configurations because it is like a pure SRAM-based architecture) and after this, the 32K(6:2) setting consumes less energy than all other 4 configurations for the 32KB cache size. The order is the same with the 32KB cache size because a large SRAM size gives more benefits during stable power. Compared to pure STT-RAM cache architecture, the 32K(6:2) setting consumes 42.91\% less energy, and the 32K(4:4) setting consumes 26.01\% less energy. Compared to pure STT-RAM cache architecture, the 32K(2:6) setting consumes 13.37\% less energy.


The above analysis concludes that pure SRAM-based architecture i,e. 16K/32K (8:0) performs better, and then 16K/32K (6:2) performs better than other configurations. However, this does not imply that the 16K/32K (0:8) and the 16K/32K (6:2) architectures are preferable because SRAM has a relatively high leakage energy than STT-RAM, whereas STT-RAM has 3x times of density than SRAM. As a result, when selecting hybrid architectures, the size of NVMs (both at cache and main memory), associativity, and energy consumption must all be considered.




\subsubsection{\textbf{Under Frequent Power failures} }

We assume frequent power failures happen for every 2 and 4 million instructions. We perform all experiments for one billion instructions in the gem5 simulator. We modeled three power failure scenarios, as shown in table \ref{tab6}. In case 1, power failures occur for every 2 million instructions. In case 2, power failures occur for every 4 million instructions. In case 3, power failures occur randomly in between 2 to 4 million instructions.

\begin{table}[htp]
\centering
\caption{Different Power Failure Scenarios}
\label{tab6}
{
\begin{tabular}{|l|l|}
\hline
\textbf{Configuration} & \textbf{Power Failure (PF) Scenario} \\ \hline
Case-1 (Proposed 2M)               & PF for every 2-Million Instructions  \\ \hline
Case-2     (Proposed 4M)            & PF for every 4-Million Instructions  \\ \hline
Case-3 (Proposed Random) & \begin{tabular}[c]{@{}l@{}}Random PF between every \\ 2 to 4-Million Instructions\end{tabular} \\ \hline
\end{tabular}%
}
\end{table}

Considering energy harvesting sources, such as piezoelectric and vibration-based sources, they extract much less energy from the surroundings. In these cases, the capacitor cannot store enough energy, resulting in frequent power failures. As a result, our proposed architecture supports these worst-case scenarios. However, existing work by Xie et al. made similar assumptions, assuming that each power failure occurs for every 500 ms.

\begin{figure}[htbp]
  \includegraphics[width= 1\linewidth]{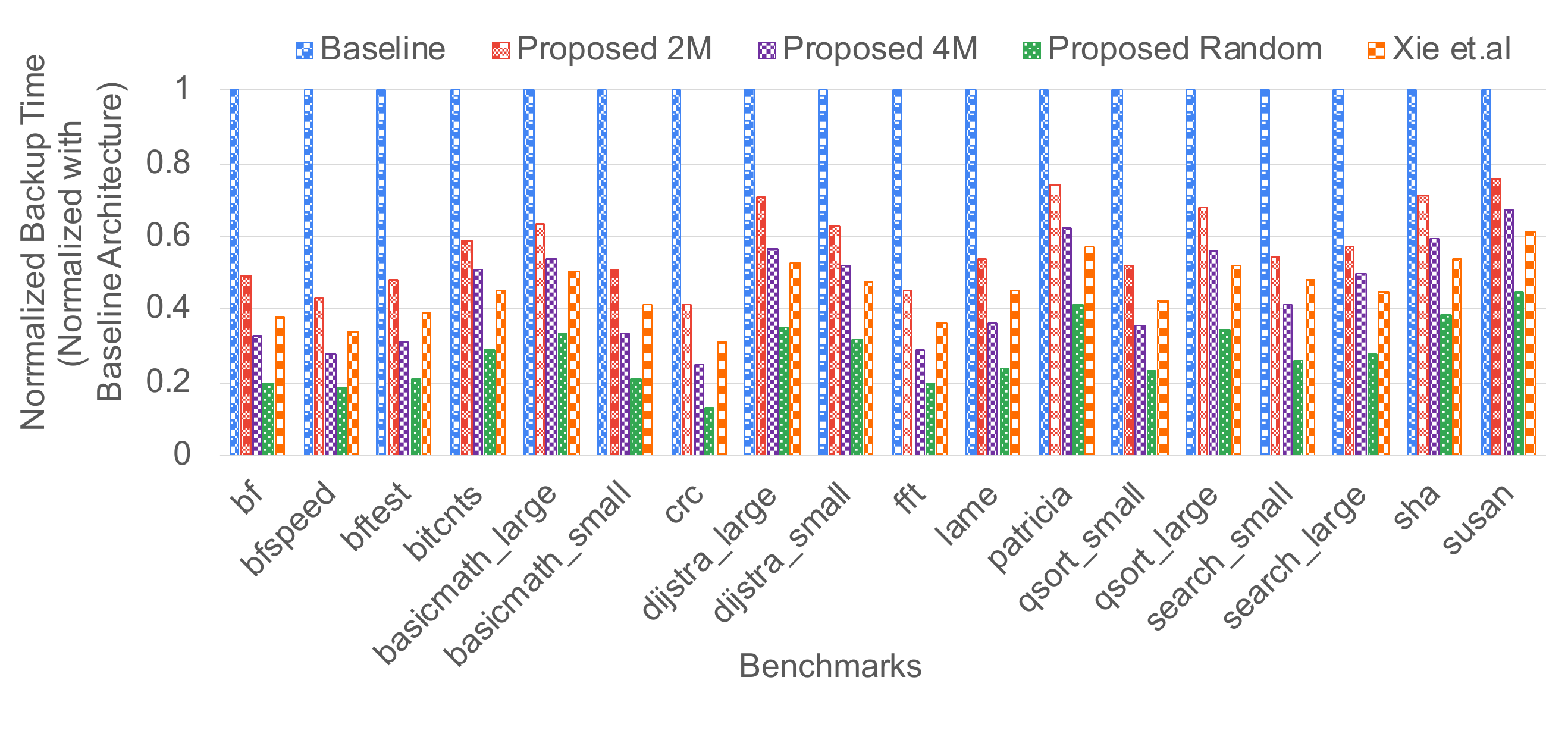}
\caption{ Backup Time }
\label{tab4}
\end{figure}

\begin{figure}[htp]
  \includegraphics[width= 1\linewidth]{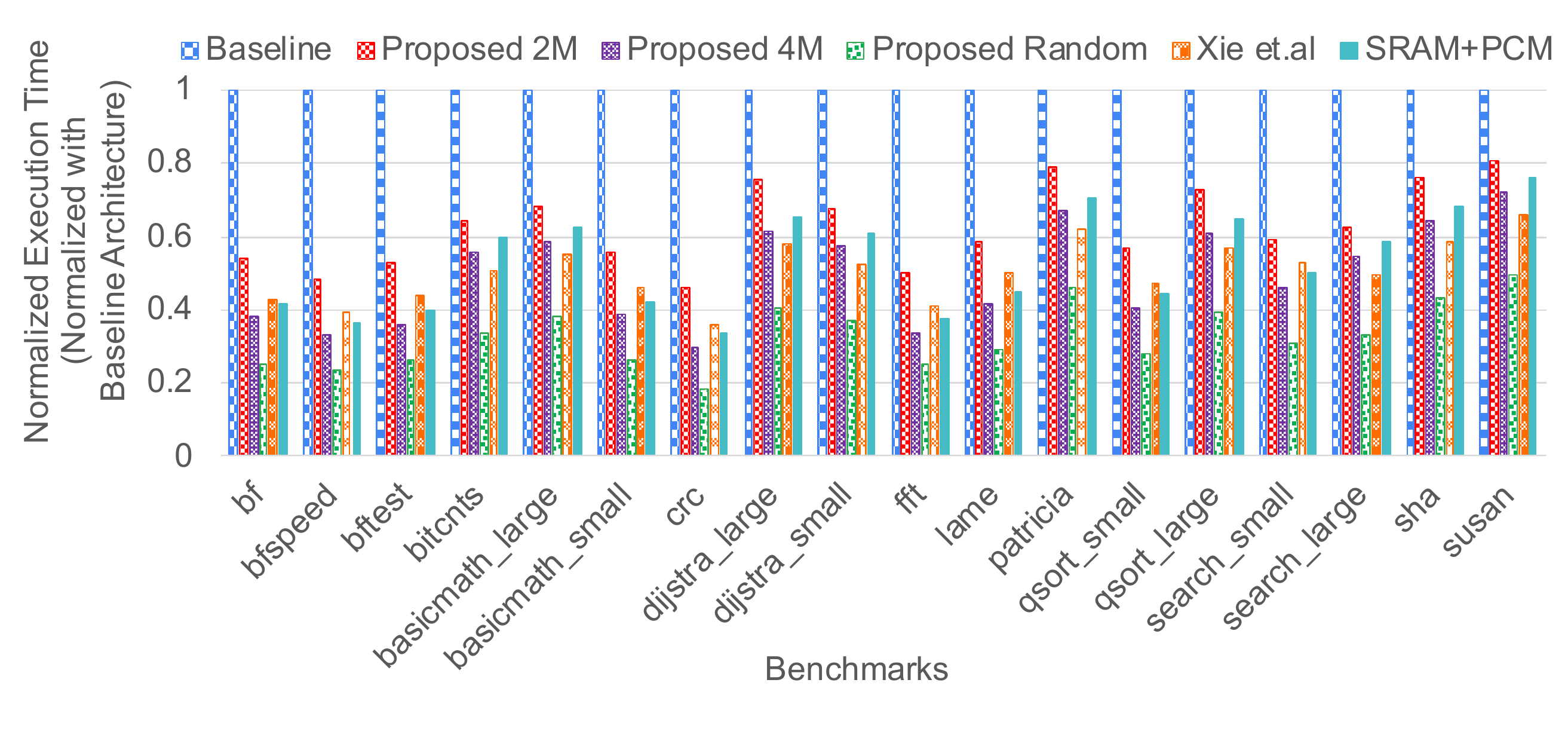}
\caption{ Comparisons between Proposed, Baseline, and Existing Architectures for Execution Time under Frequent Power Failures. }
\label{tab41}
\end{figure}

\begin{figure}[htp]
  \includegraphics[width= 1\linewidth]{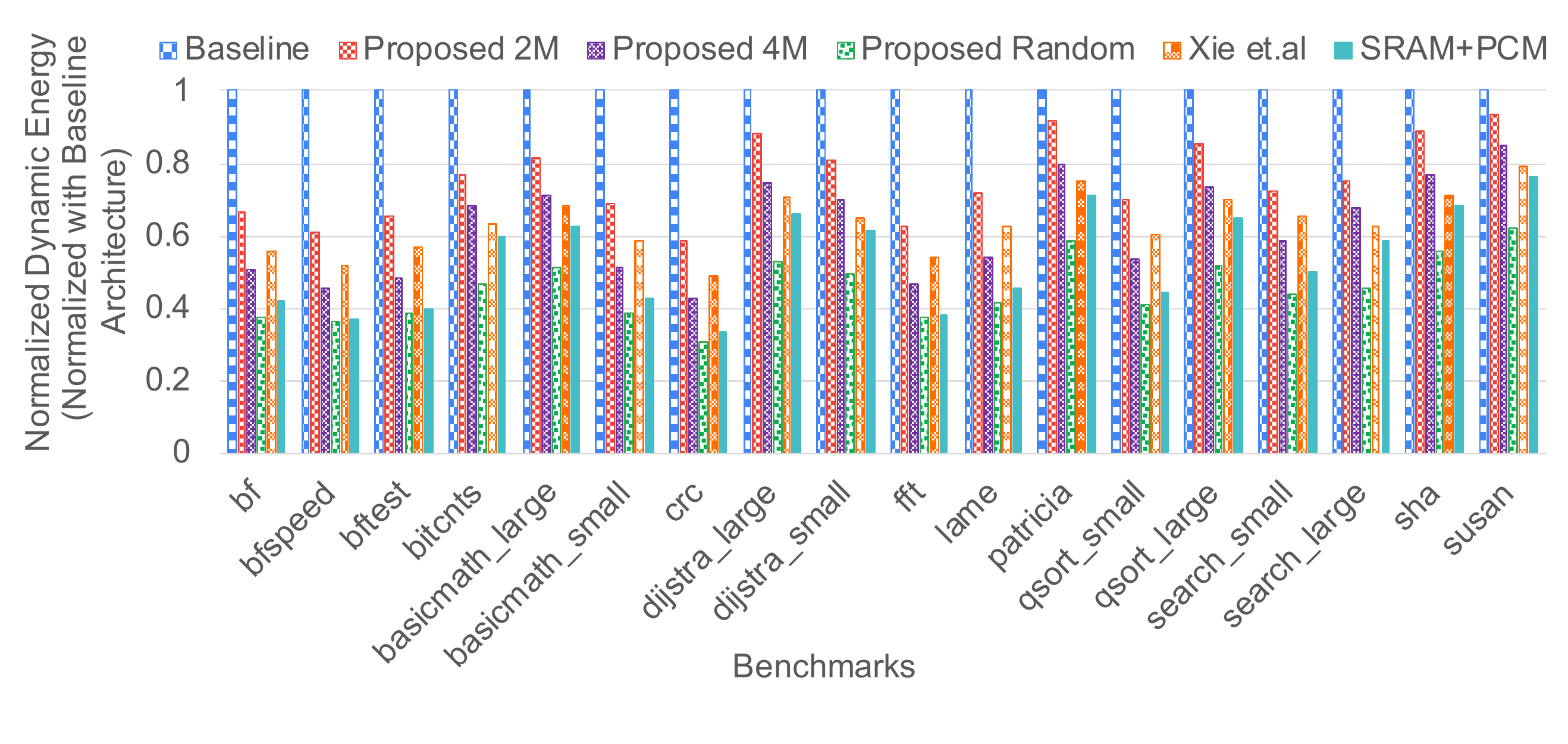}
\caption{ Comparisons between Proposed, Baseline, and Existing Architectures for Dynamic Energy Consumption under Frequent Power Failures. }
\label{tab4111}
\end{figure}

In the figures \ref{tab4}, \ref{tab41}, and \ref{tab4111}, we refer to proposed 2M with case-1, proposed 4M with case-2, and proposed random with case-3. We calculated the average backup time ($B_t$), i.e., the time required to backup all the SRAM contents to NVM. We also evaluate a random intermittent power system, where power failure occurs very often and randomly, to check $B_t$ and the efficiency of the proposed architecture. The performance and energy consumption values are normalized based on the baseline architecture.

We compare the average $B_t$ w.r.t to the baseline, as shown in figure \ref{tab4}. We also compared SRAM+PCM-based architecture to show how much performance improved during intermittent power supply. In SRAM+PCM architecture, SRAM is the L1 cache, and PCM is the main memory. We introduced a power failure randomly and a safe point for every 4 million instructions. When a power failure occurs, we back up all SRAM contents to PCM. Whenever power comes back, we start the application's execution from the nearest safe point. When we compared SRAM+PCM architecture with the proposed architecture, the proposed architecture gives better because the proposed architecture saves data at the L1 cache itself (by using STT-RAM). proposed architecture saves the re-execution time of the application and reduces the number of writes to PCM during a power failure. The performance and energy consumption values are normalized with the baseline architecture. We compare the execution time and energy consumption with the baseline architecture during these frequent power failures, as shown in figures \ref{tab41} and \ref{tab4111}.

We also compared the proposed architecture with the existing work, i.e., Xie et al. They checkpoint only selective dirty blocks from SRAM to STT-RAM during power failures. This type of checkpointing increases writes to PCM, which increases dynamic energy consumption for their architecture. Thus, the proposed architecture achieves better execution time and energy values than the existing architecture. Whenever power comes back, the proposed architecture uses blocks from STT-RAM directly. In Xie et al. work, STT-RAM consists of fewer blocks than the proposed architecture, which increases execution time in existing work.

\begin{figure}[htp]
    \centering
    \subfloat[\centering Execution Time ]{{\includegraphics[width=1\linewidth]{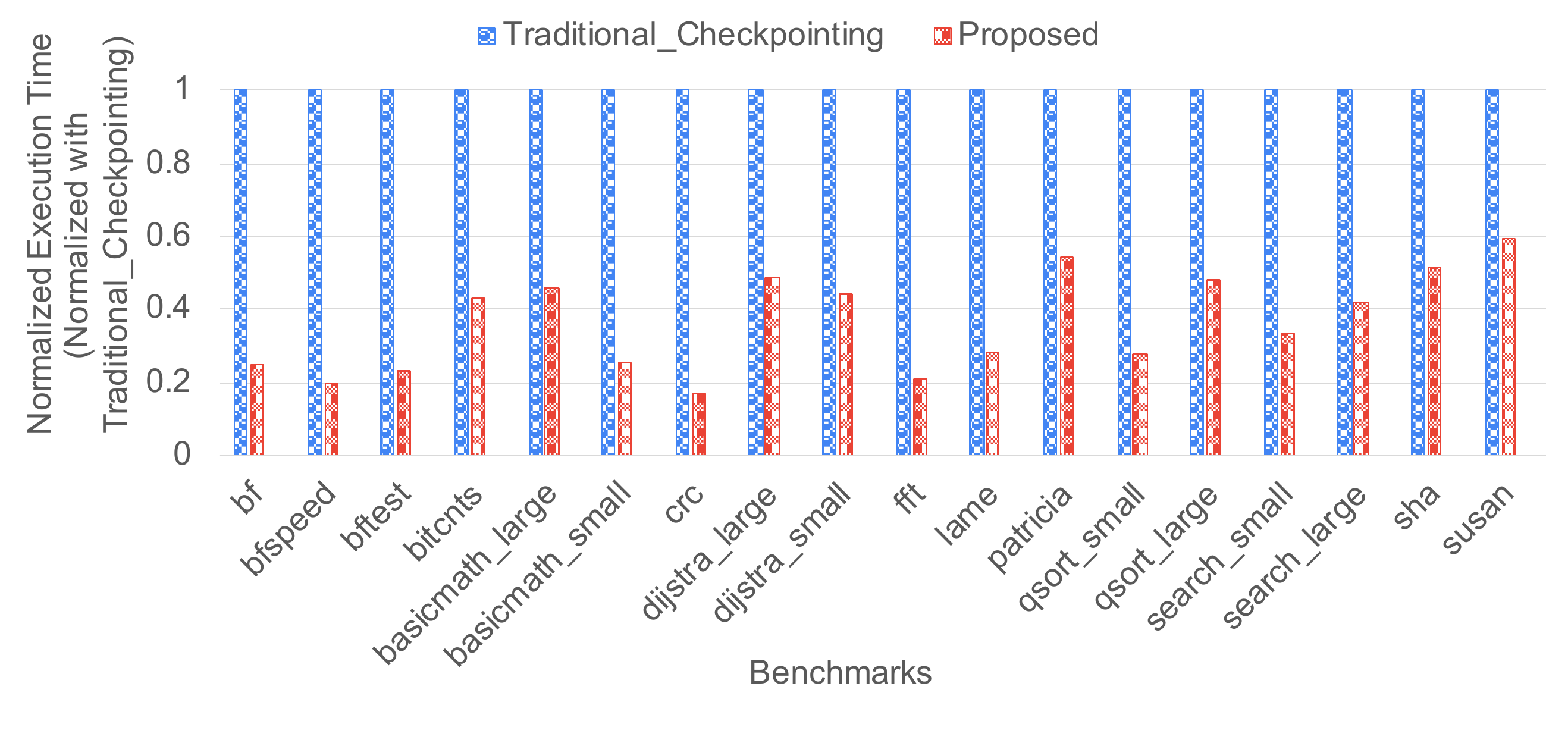} }}%
    \qquad
    \subfloat[\centering Dynamic Energy Consumption ]{{\includegraphics[width=1\linewidth]{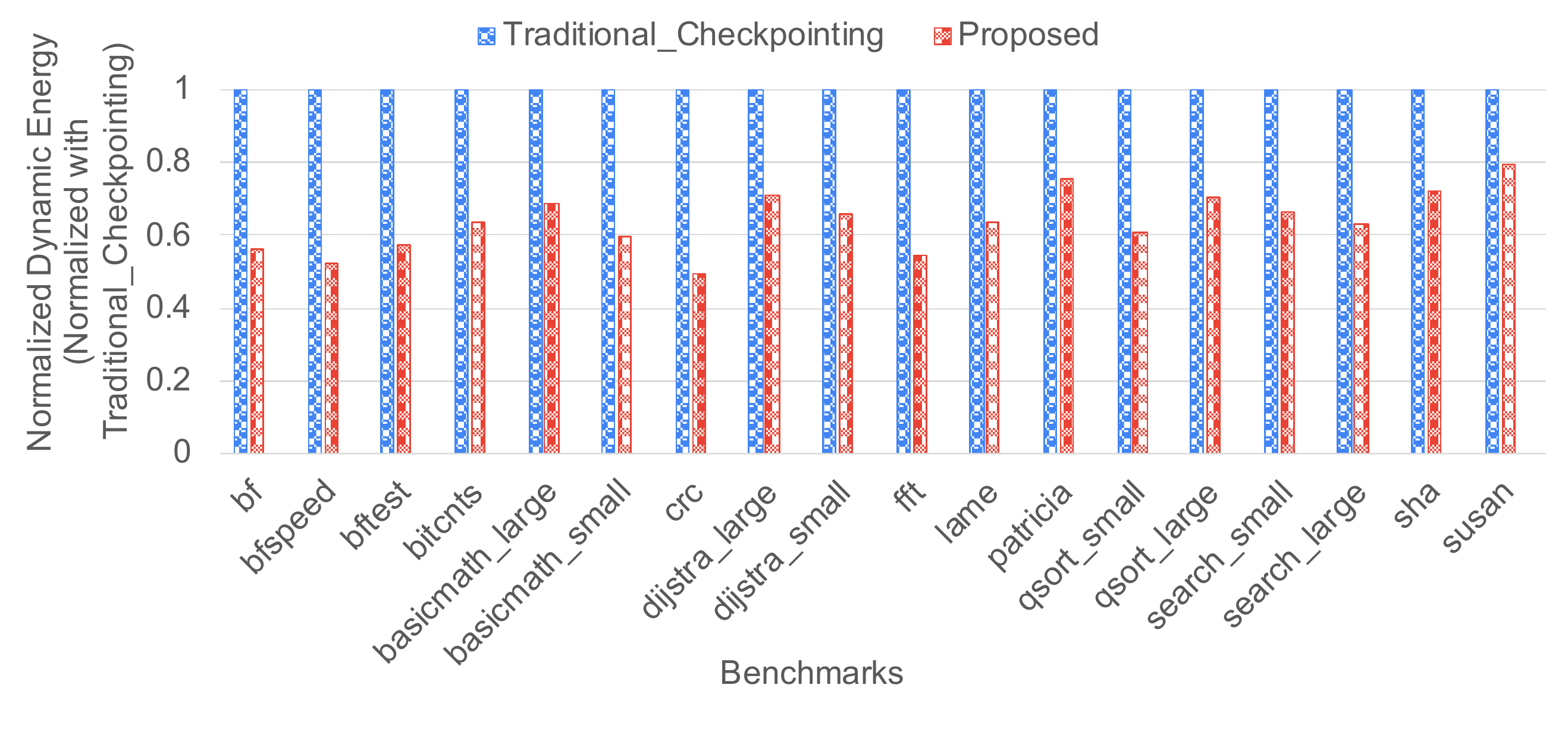} }}%
    \caption{Comparison in terms of Performance Overhead and Energy Consumption during Power Failure}%
    \label{Fig33}%
\end{figure}


We compare the traditional checkpointing approach with the proposed architecture during power failures. As earlier said, we implemented a traditional checkpointing approach by creating a safe point for every 4 million instructions. We save the program state for every 4 million instructions. We retrieve the program state from the main memory at every safe point to continue with the remaining execution of the application. For instance, if a random power failure occurs at $9^{th}$ million instruction. We re-execute the application from $4^{th}$ million instruction because the nearest safe point is at $4^{th}$ million instruction. The performance and energy consumption values are normalized based on the traditional checkpointing approach. We compared the proposed architecture with the traditional checkpointing approach, which reduces performance overhead and energy consumption by 36.10\% and 31.03\%, as shown in figure \ref{Fig33}. 




\textbf{Analysis for Different Cache Settings:} We also performed experiments by changing the system configurations like cache sizes and associativity that are different from the system configuration shown in table \ref{tab1}. We used 10 different cache settings for these experiments, as shown in table \ref{tab7}.

\begin{figure}[htp]
  \includegraphics[width= 1\linewidth]{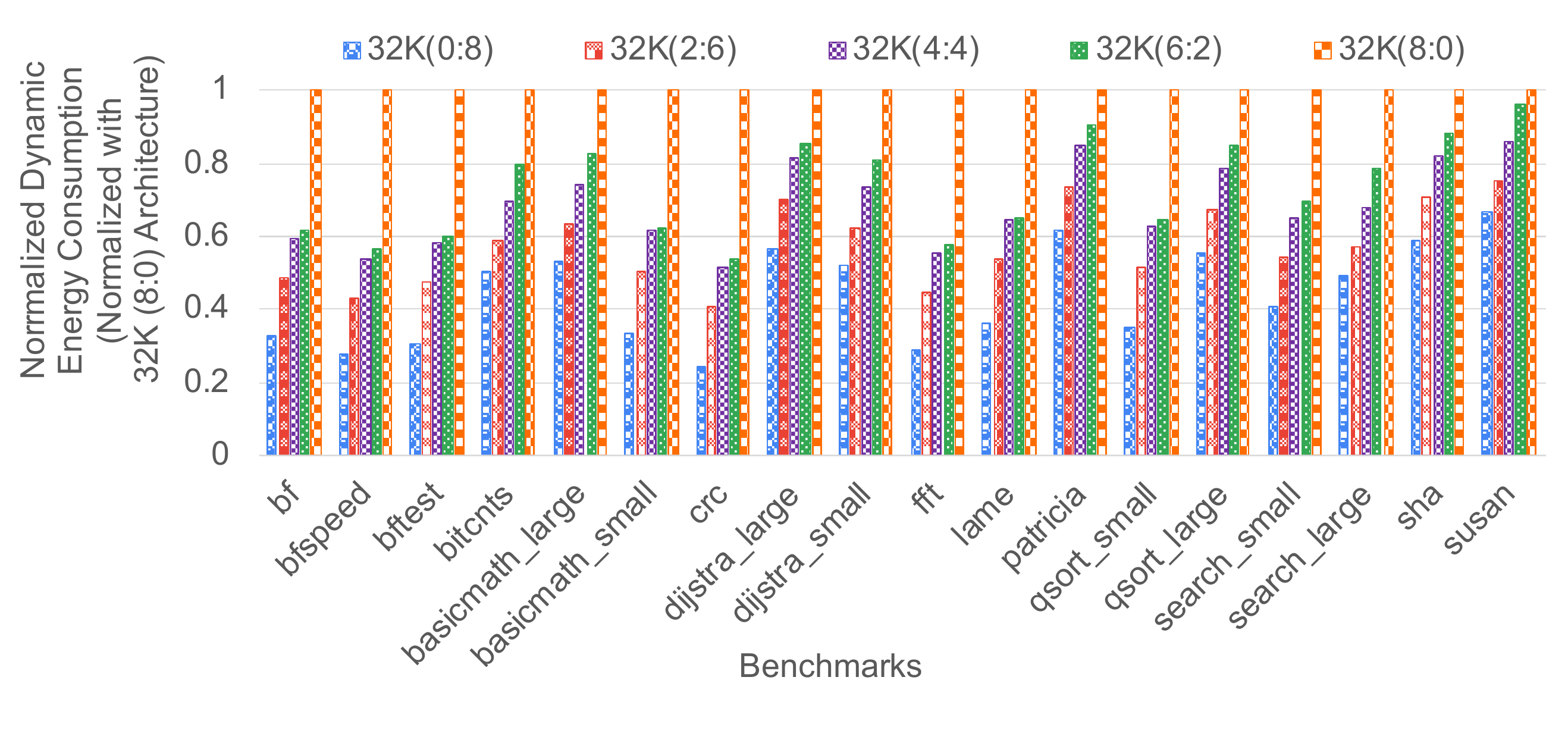}
\caption{ Dynamic Energy Consumption for Different Cache Configurations under Unstable Power, where Cache Size is 32KB }
\label{Fig3301}
\end{figure}

Here, we used two different cache sizes; one is 16 KB, and the other is 32 KB. We compared the energy consumption under an unstable power supply for the 10 configurations. We used all proposed policies and techniques in these 10 sets of configurations. In figure \ref{Fig3301}, the energy consumption of the configurations \{6-9\} is normalized based on configuration-10 (pure SRAM-based cache architecture for 32KB cache size).



We observed that for the 16KB cache size, configuration-5 consumes more energy than the other 4 configurations during an unstable power because of backing up SRAM contents to SRAM. During an unstable power supply, we observed that the configuration with more STT-RAM ways consumes less energy than others without relating to cache sizes. We observed 16K(0:8) setting consumes less energy than all other 4 configurations because it is like a pure STT-RAM-based architecture) and after this, the 16K(2:6) setting consumes less energy than all the other 3 configurations. After 16K(2:6), the 16K(4:4) setting consumes less energy than the other two configurations, i.e., 16K(6:2) and 16K(8:0). Compared to pure SRAM cache architecture, the 16K(2:6) setting consumes 16.70\% less energy, the 16K(4:4) setting consumes 12.19\% less energy. Compared to pure STT-RAM cache architecture, the 16K(6:2) setting consumes 7.11\% less energy.  

Similarly, We observed 32K(0:8) setting consumes less energy than all other 4 configurations because it is like a pure STT-RAM-based architecture) and after this, the 32K(2:6) setting consumes less energy than all other 3 configurations for the 32KB cache size. The order is the same as the 16KB cache size because a large STT-RAM size gives more benefits during unstable power, where it backup more data and reduces both backup and restore overhead. Compared to pure SRAM cache architecture, the 32K(2:6) setting consumes 21.10\% less energy, and the 32K(4:4) setting consumes 15.49\% less energy, as shown in figure \ref{Fig3301}. Compared to pure STT-RAM cache architecture, the 32K(6:2) setting consumes 9.14\% less energy, as shown in figure \ref{Fig3301}.

The above analysis concludes that pure STT-RAM-based architecture i,e. 16K/32K (0:8) performs better, and then 16K/32K (2:6) performs better than other configurations. However, this does not imply that the 16K/32K (8:0) and the 16K/32K (2:6) architectures are preferable because STT-RAM has relatively high read/write latency and consumes more dynamic energy than SRAM. Thus, we used and suggested equal partitions of SRAM and STT-RAM throughout this work, which give benefits under both stable and unstable power supplies.

\begin{figure}[htp]
  \includegraphics[width= 1\linewidth]{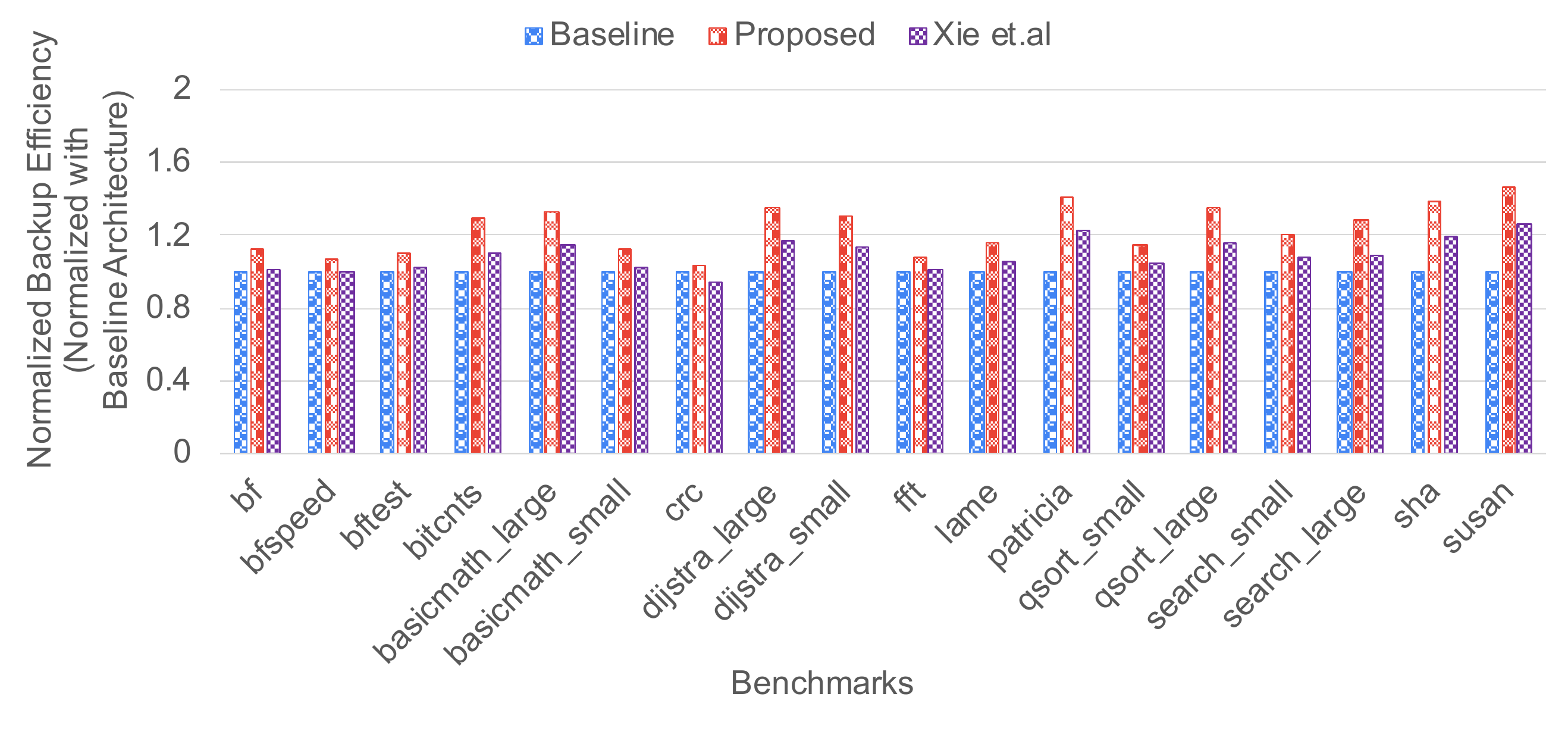}
\caption{ Comparison of Backup Efficiency ($\eta$) during Power failures}
\label{Fig321a}
\end{figure}

\begin{figure}[htp]
  \includegraphics[width= 1\linewidth]{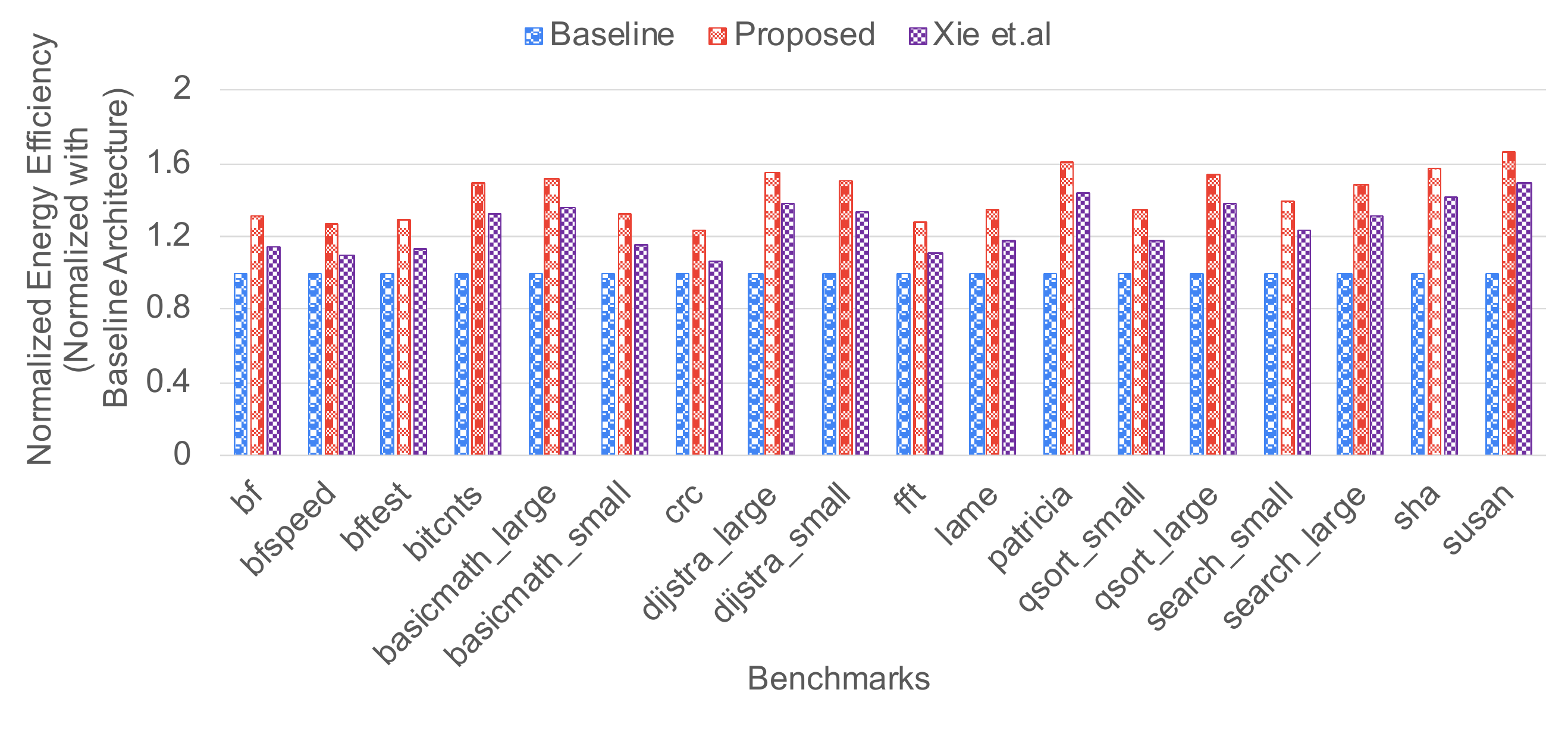}
\caption{ Comparison of Energy Efficiency ($\theta$) during Power failures}
\label{Fig321b}
\end{figure}


As shown in figures \ref{Fig321a} and \ref{Fig321b}, we performed experiments to analyze the backup efficiency ($\eta$) and energy efficiency ($\theta$) for both proposed and existing architectures. The $\eta$ and $\theta$ values are normalized with the baseline architecture. Our proposed architecture improves $\eta$ by 32.52\% and $\theta$ by 43.41\% because of the proposed backup strategy. The other reason for the improvement in both $\eta$ and $\theta$ is a reduction in both $E_{backup}$ and $B_t$.

Lastly, as we discussed SRAM+PCM architecture, there is a safe point for every 4 million instructions. Whenever power failure occurs, we save the state in PCM. This type of backup policy increases writes to PCM. Whenever power comes back, the restore procedure increases the number of accesses from PCM to the SRAM cache. In a hybrid cache, STT-RAM saves some blocks so that PCM observes fewer writes, and the restore takes lesser accesses from PCM. We evaluated the 32KB SRAM cache and hybrid cache (16KB SRAM+16KB STT-RAM) to check static power. We have seen the proposed architecture has a 17.02\% improvement in static power compared to 32KB SRAM+PCM architecture.

\section{Conclusions} \label{p6}


The proposed architecture is a promising HCA for IoT embedded systems. The proposed architecture is beneficial for IoT applications, where power failures are frequently unpredictable. Because of its high write latency and energy consumption, NVM introduces overhead in hybrid caches. We proposed an efficient prediction-based placement policy and an intelligent migration policy that efficiently uses SRAM and STT-RAM. We reduce the number of writes to STT-RAM by effectively using the proposed prediction table. In comparison to the baseline architecture, the proposed architecture reduces STT-RAM writes from 63.35\% to 35.93\%. As a result, our energy consumption and execution time are reduced.

We compared the proposed architecture to state-of-the-art and baseline architectures. proposed improves energy and backup efficiency. We proposed a backup strategy to ensure the efficient backup of the program state. During a power failure, the proposed backup strategy helps to recognize important blocks and migrate them to the STT-RAM cache. When compared to baseline and existing architectures, proposed requires less backup time. When power comes back, we use STT-RAM contents without any restoration procedure.

\bibliography{main}


\end{document}